\documentclass[journal]{IEEEtran}

\usepackage{multirow}
\usepackage{graphicx}
\usepackage[table,xcdraw]{xcolor}
\usepackage[cmex10]{amsmath}
\usepackage{cite}
\usepackage{amsmath,amssymb,amsfonts}
\usepackage{siunitx}    
\usepackage{ragged2e}
\usepackage{caption} 
\usepackage{booktabs}
\usepackage{hyperref}
\usepackage{booktabs}
\usepackage{mwe}
\usepackage{booktabs, makecell, multirow, tabularx}
\setcellgapes{2pt}
\newcolumntype{L}[1]{>{\RaggedRight\hspace{0pt}%
                     \hsize=#1\hsize}X}
\usepackage{lipsum}
\usepackage{atbegshi}
\AtBeginDocument{\AtBeginShipoutNext{\AtBeginShipoutDiscard}}

\begin{document}
\bstctlcite{IEEEexample:BSTcontrol}
    \title{Multifunction full-space graphene-assisted metasurface}
  \author{Parsa~Farzin,
      Amir saman~Nooramin,
      and Mohammad~Soleimani}

  \thanks{School of Electrical Engineering, Iran University of Science and Technology, Tehran, 1684613114, Iran}


\maketitle

\begin{abstract}
In recent years, there has been notable advancement in programmable metasurfaces, primarily attributed to their cost-effectiveness and capacity to manipulate electromagnetic (EM) waves. Nevertheless, a significant limitation of numerous available metasurfaces is their capability to influence wavefronts only in reflection mode or transmission mode, thus catering to only half of the spatial coverage. To the best of our knowledge and for the first time, a novel graphene-assisted reprogrammable metasurface that offers the unprecedented capability to independently and concurrently manipulate EM waves within both half-spaces has been introduced in the THz frequency band. This intelligent programmable metasurface achieves wavefront control in reflection mode, transmission mode, and the concurrent reflection-transmission mode, all within the same polarization and frequency channel. The meta-atom is constructed with three graphene sections, enabling straightforward modification of wave behavior by adjusting the chemical potential distribution within each graphene segment via an external electronic source. Beyond real-time control of reflection and transmission modes, this approach also empowers the manipulation of wavefronts by governing the phases of these modes. The proposed functionalities encompass various programmable modes, including single and dual beam control in reflection mode, dual beam control in transmission mode, simultaneous control of direct transmission alongside two beams in reflection mode, and vice versa, and controlling a single beam in reflection mode and direct propagation and two beams in transmission mode simultaneously. Furthermore, we extended our exploration beyond wavefront manipulation in individual or continuous reflection and transmission modes. In fact, the structure possesses the capability to transform the proposed metasurface into an absorber and polarizer, as well. By changing the chemical potential of each graphene segment, it can selectively absorb incident waves at different frequencies. Also, by manipulating the incident wave polarization, it can convert linear polarization to circular polarization in both reflection and transmission modes. The proposed metasurface is expected to be reprogrammable due to wavefront manipulation in both half-spaces separately and continuously as well as absorption frequency control and convert polarization for various applications such as imaging systems, encryption, remote sensing, miniaturized systems, and next-generation wireless intelligent communications.
\end{abstract}


%
\IEEEpeerreviewmaketitle


\section{Introduction}

\IEEEPARstart{I}{n} recent years, the scope of Terahertz (THz) science and technologies has reached maturity and attracted massive attention due to their potential applications in biomedicine, wireless communication, and imaging \cite{1,2,3,4}. Efforts to fully harness the benefits of THz waves have spurred significant advancements in modern and multifunctional THz devices. For a long time, the flexible control of electromagnetic (EM) waves has been a crucial subject. Metamaterials have emerged as a concise and efficient method to manipulate EM waves, leading to the rapid expansion of metamaterial technology, which now offers feasible high-technological THz platforms. These advancements have showcased remarkable progress in recent times \cite{5,6}. 

In the past decade, metamaterials have attracted much attention in scientific research and engineering fields due to their unusual and tunable EM properties that are not attainable in natural materials. Metamaterials, typically composed of artificially periodic or quasi-periodic structures with sub-wavelength scales, offer a new design strategy for functional materials, leading to exotic phenomena and applications such as negative refraction \cite{7}, zero refractive index \cite{8}, perfect absorption \cite{9}, and cloaking \cite{10}. However, metamaterials encounter significant practical constraints due to lossy characteristics, the strong dispersion of resonant responses, and the manufacturing complexities of three-dimensional bulky structures \cite{11}. As an alternative, two- dimensional (2D) counterparts of metamaterials, known as metasurfaces (MSs), have been intensively investigated due to their promising advantages, including compactness, low cost, high surface integrity, and ease of fabrication \cite{12}. They thereby overcome the challenges encountered by their 3D counterparts \cite{13}. As an emerging platform, MSs are inspiring a flourishing of research interest with exceptional manipulation of EM wave amplitude \cite{14,15}, phase \cite{16,17,18,19}, and polarization \cite{20,21}. In this regard, numerous MSs have been developed to achieve extraordinary applications, such as meta-lenses \cite{22}, orbital angular momentum generation (OAM) \cite{23}, holography \cite{24,25}, and cloaking \cite{26}. As one of the most promising candidates, T. Cui et al. have recently introduced the concept of digital MSs, revolutionizing the field by linking the physical and digital worlds. This innovative approach enables a fresh perspective on MSs from the standpoint of information science \cite{27}. Unlike conventional MSs, coding MSs with well-defined elements can be digitalized and programmed using a field-programmable gate array (FPGA). By controlling sequences of digital coding states "0" and "1" with opposite-phase responses (0° and 180°), the MSs allow for the manipulation of EM waves with various functionalities \cite{28}. Moreover, this coding and programmable MSs facilitate dynamic and real-time control of EM waves, offering the extraordinary potential to establish dynamic MSs. The emergence of multifunctional MSs structures has attracted considerable attention in recent years. Unlike previous single-functional MS, these versatile MSs can now achieve a diverse range of electromagnetic functions simultaneously \cite{29}. Despite the great achievements attained so far, we note that the wave manipulation capabilities of multifunctional MSs have been much less explored \cite{30}. Moreover, the manipulation of EM waves by these MSs is usually limited to half-space, and most devices can only operate in reflection \cite{31,32} or transmission mode \cite{33}. Consequently, they can only manipulate the reflected or transmitted wave, leaving half of the EM space unutilized \cite{34}. This limitation significantly hinders their potential applications. While some MSs studies have proposed to manipulate the wavefront of both the reflected and transmitted waves in the full space. In \cite{35}, Pan et al., MS designed in the microwave band achieves remarkable EM wave control for both reflection and transmission modes through dimension adjustments, effectively operating in two frequency bands: reflecting x-polarization and transmitting y-polarization in the $f_1$ frequency band, also transmitting x-polarization in the $f_2$ frequency band. Phase control is also achieved by changing dimensions, enabling precise manipulation of the EM waves. The reflected and transmitted waves are manipulated using orthogonal grating-wire layers, thus allowing for control over the EM waves. In another study, described in \cite{36}, Wu et al., researchers demonstrated MSs operating in the microwave band that effectively controls y-polarized waves for both reflected and transmitted wave modes by adjusting the bias of the diode pin. Notably, phase control is achieved through dimension changes of the patch, enabling wave manipulation in both reflection and transmission modes. Finally, in \cite{37,38}, T. Cui et al. and Bao et al., proposed a structure operating in the microwave band, which utilizes pin diodes to control not only the reflected and transmitted EM waves but also the phase. This design allows real-time manipulation of waves in full space. However, as the frequency increases from microwave to terahertz, the availability of PIN diodes becomes limited, making their use impractical \cite{39}. Consequently, achieving programmable terahertz metamaterials that match the performance of existing microwave materials poses a significant challenge. Researchers to overcome this limitation have turned to active materials such as graphene \cite{40,41}, vanadium dioxide ($VO_2$) \cite{42,43}, InSb \cite{44}, and liquid crystals \cite{45} to manipulate terahertz EM waves. However, achieving control of reflected and transmitted waves in MSs is not limited to the microwave band, and efforts have also been made in the THz band. In \cite{46}, Dong et al., the proposed structure achieves control of the reflection and transmission of EM waves in two different frequency bands through the use of InSb temperature changes. Additionally, phase control is obtained by adjusting the dimensions of the structure. To the best of the author’s knowledge, all the research in the THz band that has been done on the real-time manipulation of reflected and transmitted waves in an MS involves at least one task of controlling the amplitude (reflection or transmission) or the phase of the EM wave by changing the dimensions \cite{47,48,49,50,51}. Therefore, in all of these strategies, MSs are designed for a specific application, and once fabricated, their performance remains constant.

Graphene, a flat monolayer of carbon atoms organized in a two-dimensional (2D) honeycomb-like lattice, has garnered significant attention worldwide due to its exceptional electrical and mechanical properties and the design freedom it offers \cite{52,53}. The ability to arbitrarily control the electrical properties of graphene through external biasing has opened up possibilities for the development of radically different photonic devices \cite{54}. By harnessing this distinctive property of graphene, it becomes possible to obtain distinct amplitude and phase responses, leading to diverse capabilities, including manipulating wavefront \cite{55}, polarization converters \cite{56,57,58}, and tunable absorption \cite{59}.

Although we have recently introduced a graphene-based metasurface that enables control over reflection, transmission, absorption, and polarization conversion for both reflection and transmission modes, our current design is restricted by its lack of phase control, limiting its manipulation of the wavefront across two half-spaces \cite{60}. To address these constraints, For the first time to the best of our knowledge in the THz frequency band, we introduce an intelligent reprogrammable graphene-assisted metasurface that holds the capability to independently and concurrently manipulate EM wavefront across both half-spaces in real-time at the same polarization and frequency channel. The meta-atom is constructed with three graphene sections, enabling straightforward modification of wave behavior by adjusting the chemical potential distribution within each graphene segment via an external electronic source by FPGA. Through the utilization of two graphene sections within the meta-atom, independent or simultaneous control of reflection, transmission, and phase adjustments can be achieved across the entire space in real-time. By using two graphene sections integrated in the meta-atom, it becomes feasible to independently or concurrently manipulate reflection, transmission, and phase characteristics throughout the entire space in real-time. To assess the efficacy of the proposed metasurface, we explore diverse functionalities in both reflection and transmission modes, both individually and concurrently. These include single and dual beam control in reflection mode, dual beam control in transmission mode, simultaneous management of direct transmission alongside two beams in reflection mode, and vice versa, and dynamic control of a single beam in reflection mode along with two beams in transmission mode, all within the same polarization and frequency channel. Moreover, our investigation extends beyond wavefront manipulation solely in separate or continuous reflection and transmission modes. the proposed metasurface has the added potential to be transformed into an absorber and polarizer. Through precise adjustment of the chemical potential within each graphene segment, we can selectively absorb incident waves at specific frequencies and effectuate the conversion of linear to circular polarization across both reflection and transmission modes. We firmly believe that the introduced metasurface elevates the landscape of intelligent multifunctional metasurfaces and carries substantial potential across a range of applications, including optical communication, imaging systems, and next-generation wireless intelligent communications.

\section{Materials and Methods}
\subsection{Complex Graphene's surface conductivity}

Graphene, a 2D material, consists of carbon atoms arranged in a hexagonal lattice, garnering widespread attention in the last decade as a zero-gap semiconductor with unique electrical, thermal, optical, and mechanical properties. Furthermore, owing to its boundary conditions, graphene's extreme sensitivity to external stimuli makes it an excellent candidate for manipulating THz waves. Through the Hall Effect, the AC conductivity of graphene can be tuned using external electrostatic or magnetostatic bias. Moreover, due to its mono-atomic structure, graphene can be locally represented by a complex surface conductivity tensor \cite{61,62}:

\begin{equation}
   \sigma(\omega_c, \mu(E_0 ), \Gamma, T, B_0 )= \hat{x}\hat{x}\sigma_{xx} + \hat{x}\hat{y}\sigma_{xy} + \hat{y}\hat{x}\sigma_{yx} + \hat{y}\hat{y}\sigma_{yy}
   \label{eq1}
\end{equation}

where $\omega$ is the radian frequency, $\mu_c$ is the chemical potential, $\Gamma=1⁄2\tau$ is the phenomenological scattering rate with $\tau$ being the electron-phonon relaxation time. T is the room temperature, $E_0$ and $B_0$ are electrostatic and magnetostatic bias fields, respectively. Without magnetostatic bias, the off-diagonal components of the surface conductivity tensor become vanish, resulting in the notice of isotropic behaviors in the graphene monolayer \cite{61,62}. The graphene monolayer can be represented electrically as an infinitely thin conducting layer with a surface resistance $R_g$  characterized by a complex-valued conductivity surface. According to the Kubo formula, the complex surface conductivity of graphene can be expressed with the help of interband and intraband transitions \cite{61,63}.

\begin{equation}
   R_g = 1 ⁄ \sigma_s = 1 ⁄(\sigma^s_{intra}+\sigma^s_{inter}) 
   \label{eq2}
\end{equation}
\begin{equation}
    \sigma^s_{intraband} (\omega) = -\frac{ie^2 k_B T}{\pi \hbar^2 (\omega-i/\tau)}
    [\frac{\mu_c}{k_B T} + 2ln(e^\frac{-\mu_c}{k_B T}+1)]
    \label{eq3}
\end{equation}

\begin{equation}
    \sigma^s_{intraband} (\omega) = -\frac{ie^2}{4\pi \hbar} ln[\frac{2|\mu_c |-(\omega-i⁄\tau)\hbar}{2|\mu_c |+(\omega-i⁄\tau)\hbar}]
   \label{eq4}
\end{equation}

Here, e is electron charge, $\hbar$ represents the reduced Plack’s constant, $\mu_c$  is the chemical potential, $k_B$ is the Boltzmann’s constant. At room temperature and low terahertz frequency, interband transitions can be neglected due to the Pauli Exclusion Principle, as the photon energy is $\hbar \omega \ll E_f$ \cite{64}. In our simulations, we considered two parameters of temperature and relaxation time equal to T= 300 K and $\tau$ =0.1 and 1 ps (for different section), respectively, which are kept constant throughout this study. Additionally, the relative permittivity of graphene layers is expressed as $\epsilon_rG = 1 + \sigma_s⁄(j\omega\sigma_0 \Delta)$, where $\Delta$ denotes the ultra-thin thickness of the graphene layer. It is worth mentioning that the surface conductivity of the graphene layer can be manipulated through an external electrical bias.
By changing the chemical potential of graphene, the properties of graphene can be tuned between dielectric and conductive states. Increasing the chemical potential of graphene causes it to change from dielectric to conductor and vice versa \cite{65}. As a result, it enables real-time dynamic switching for all the different functions considered for the MS designed can be changed through an external electrical bias in this paper. Detailed information can be found in Supplementary Information A. 

\begin{figure*}
    \centering
    \includegraphics[scale=.125]{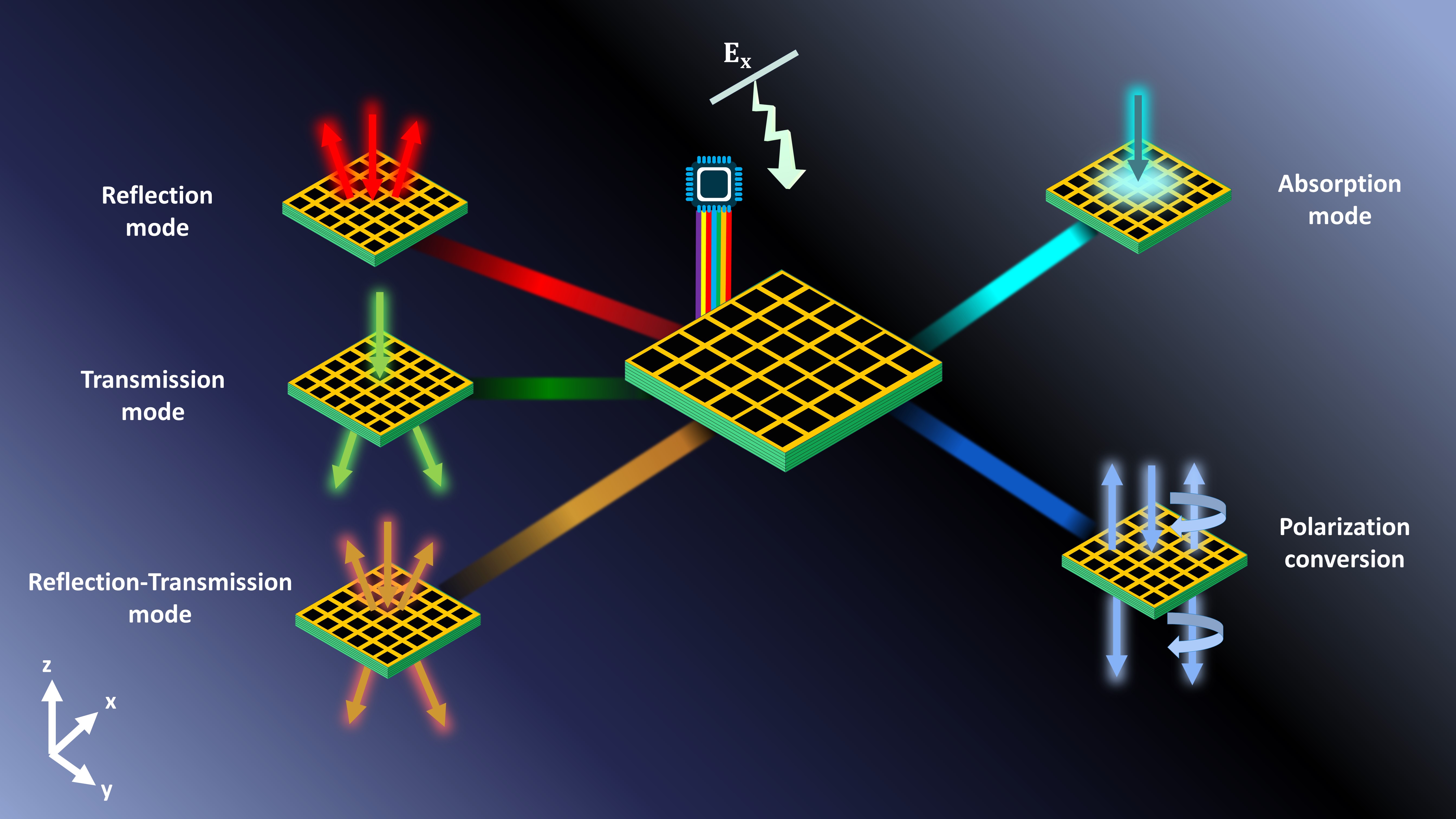}
    \caption{Imagine a versatile and programmable metasurface concept, composed of an array of metaatoms capable of dynamically altering their operational states to serve different functions. This intelligent and adaptable metasurface can seamlessly transition between reflection, transmission, simultaneous reflection-transmission, absorption, and polarization conversion modes, all achieved through the independent modulation of three distinct graphene sections by an integrated external FPGA-based controller. Through the utilization of unique preconfigured coding sequences within each of graphene sections, this intelligent metasurface can swiftly execute a diverse range of functions in real-time.}
    \label{FIG:1}
\end{figure*}
	
\subsection{Design Principle}
\begin{figure}
    \includegraphics[scale=0.35]{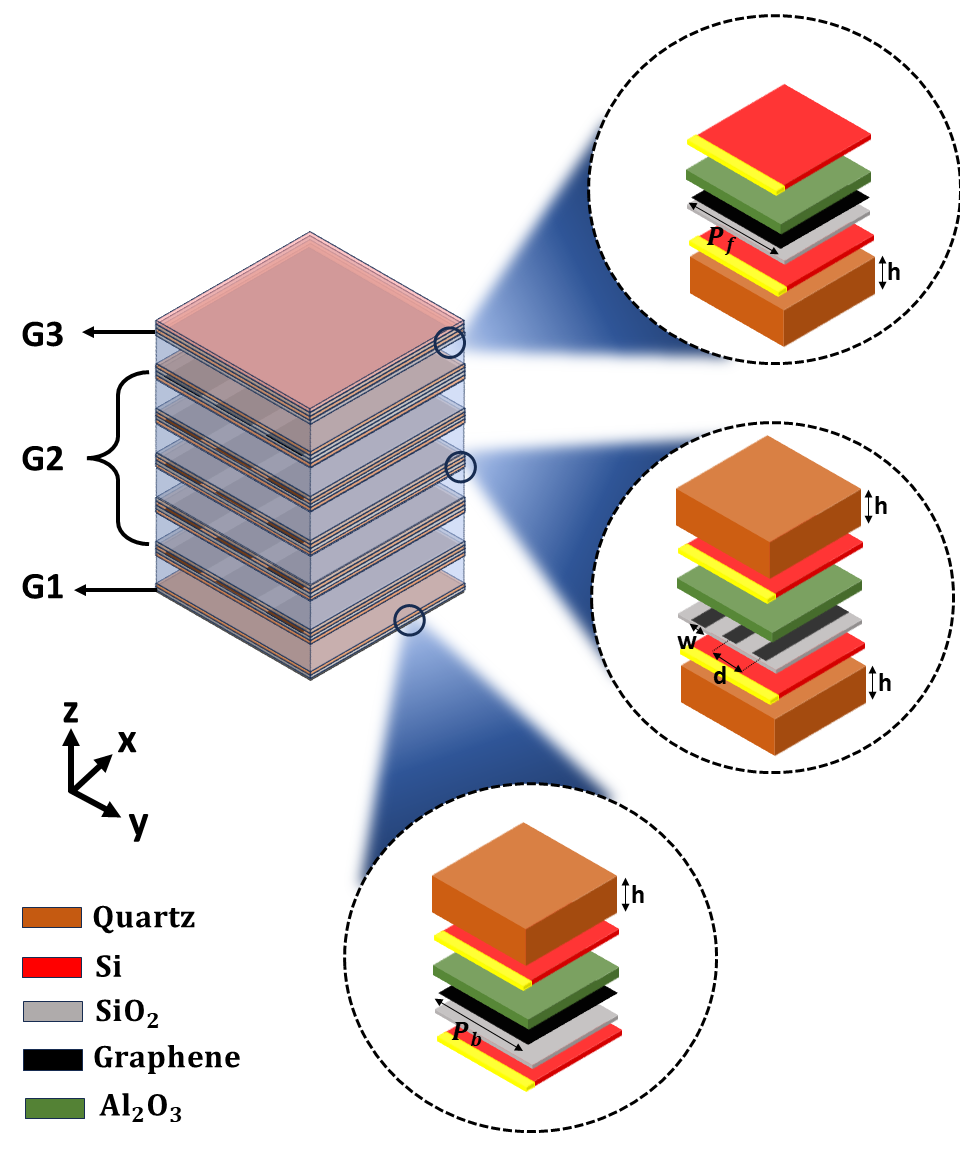}
    \caption{Programmable meta-atom geometry. The meta-atom comprises three sections of graphene denoted as G1, G2, and G3. Each section is constructed with a layered structure of Si/SiO2/Graphene/Al2O3/Si.}
    \label{FIG:2}
\end{figure}

Currently, most reconfigurable full-space metasurfaces are limited to dynamic control of only one EM property, (such as phases or reflection-transmission mode of operation). In addition, achieving full-space multifunctional capabilities with these metasurfaces are also limited to different frequency/polarization channels. One of the major significant drawbacks of single-layer metasurfaces is their limited interactions with incident EM waves, resulting in restricted EM responses \cite{66}. Thus, the potentially restricted phase modulation observed in a single graphene layer can be overcome through the implementation of a stacked structure. In this configuration, each layer effectively contributes to a wider range of phase tuning capabilities \cite{67,68}. Accordingly, several layers are needed to achieve a metasurface with different functions. To achieve real-time manipulation of the wavefront in full space, within the same frequency and polarization channel, a cascaded multilayer structure integrated with graphene is utilized. By dynamically adjusting the chemical potential of graphene, we can effectively control the transmission-reflection operation mode and phase responses in real-time. In addition to controlling the reflection, transmission, and phase of the EM wave, it can also behave as an absorber with resonance frequency shift. Furthermore, it exhibits the unique ability to convert linear to circular polarization for both reflection and transmission modes. What sets this structure apart is that all of these functions are performed in real-time, facilitated by the integration of graphene.
Figure\ref{FIG:1} shows a general schematic of the reprogrammable metasurface designed for the manipulation of EM waves on both sides of space. The proposed metasurface is composed of three distinct sections of graphene, each individually assigned to control specific functions. These three sections include reflection-transmission control section, phase control section, and absorber section. As shown in Fig. \ref{FIG:1}, By applying the appropriate bias voltage to each of the graphene parts by field-programmable gate array (FPGA), the metasurface can flexibly switch between reflection mode, transmission mode, simultaneous control mode of reflection and transmission symmetrically and asymmetrically, absorption mode, and polarization conversion mode. The functionalities of the metasurface can be dynamically controlled in real-time by adjusting the chemical potential of graphene. This real-time manipulation of the full-space occurs within the same frequency band and polarization, offering versatile and instantaneous control over the metasurface's operations. The proposed intelligent metasurface can control the reflection and/or transmission along with the manipulation of the wavefront by changing the chemical potential of the graphene layers in the backward half-space and/or forward half-space. Amplitude control is achieved by electric bias manipulation of the reflection-transmission layer, while wavefront manipulation is carried out through the phase control layer. It is worth noting that by placing reflective and transmission meta-atoms adjacently with appropriate and similar distribution, the metasurface can simultaneously control both reflection and transmission. Furthermore, by adjusting the chemical potential of the phase control layer, we can manipulate the wavefront in the forward and backward wave scatterings in the same or different ways. Controlling each characteristic of the EM wave through the design of separate layers in a metasurface not only simplifies the bias network but also grants simultaneous access to different functions. As depicted in Fig.\ref{FIG:2}, The meta-atom design includes three key components: graphene, quartz, and floating gates. These components are structured into three distinct parts: the initial square-shaped graphene section (G1) with a length of $P_b$=49 µm, the subsequent multi-layered graphene segment (G2) comprising w=5 µm wide ribbons sandwiched between quartz and floating gates, and the final square graphene section (G3) with a length of $P_f$=45 µm. The layers within G2 consist of three graphene ribbons spaced at d=10 µm intervals along the y-axis. The structure further comprises six layers of quartz, each with a uniform thickness of h=12.5 µm and possessing a relative permittivity ($\epsilon_r$) of 3.75 and a loss tangent (tand) of 0.0004. The periodicity of each meta-atom is P=50 µm (The effect of the number of layers on the amplitude and phase of the reflected and transmitted wave is given in Supplementary Information B.). Additionally, the floating gate is composed of polycrystalline silicon, alumina, and graphene. This combination enables the modulation of charge density within the graphene component, thus providing tunability to the structure. 
Further details are available in supplementary material section A.the metasurface is divided into three distinct parts, and each graphene section serves a specific function. The first part, labeled G1 and positioned at the structure's end, controls reflection transmission. In the middle of the metasurface, denoted as G2, the second part is devoted to precise phase control, utilizing multiple layers of graphene. Finally, the third part, denoted by G3 and positioned at the structure's beginning, functions as an efficient absorber. In the first section (G1), a graphene sheet is employed to control reflection transmission. By increasing (decreasing) the chemical potential of graphene, its properties can be changed from a dielectric to a conductor (or vice versa), allowing for precise control over the incident wave [60]. When graphene is in the conductor state, the structure exhibits a reflective mode. Conversely, by reducing the external electric bias and converting graphene to a dielectric, the structure switches to a transmission state. The second section (G2) utilizes five layers of graphene nanoribbons to control the phase of reflected and transmitted waves. In this section, the chemical potential of the layers changes simultaneously, yielding different phase responses. By applying appropriate chemical potentials, 1-bit, and 2-bit phase shifts can be achieved in reflection and transmission modes, respectively. In the final section (G3), the metasurface utilizes a square graphene, slightly smaller than the meta-atom dimensions. Through changing the external electrical bias G3, the structure transitions from the reflection-transmission state to the absorption state. The graphene layers employed in this paper have the same dimensions, but have different chemical potentials. 

For the purpose of modifying the charge density within each graphene layer, we employ floating gate structures, as depicted in Fig.\ref{FIG:2}. These structures consist of Si, SiO2, Al2O3, and single-layer graphene. Upon applying a positive bias voltage to the upper Si layer, electrons from the lower Si layer can tunnel through the SiO2 layer and become captured by the graphene, resulting in a heightened charge density within the graphene layer. Conversely, upon applying a reverse bias voltage to the upper Si layer, electrons from the graphene layer can tunnel through the SiO2 and become captured by the lower Si layer, leading to a decrease in the charge density of the graphene layer \cite{69,70}. Furthermore, thanks to the electrical isolation of the graphene layer from the Si layers, once the external bias voltage is removed, the charge density of the graphene can remain stable over an extended duration. Consequently, there is no requirement for any additional force to maintain the constancy of the graphene charge density. It's worth emphasizing that electron tunneling is exclusively achievable through the SiO2 layer; tunneling through the Al2O3 layer is rendered impractical due to the substantial thickness of Al2O3 compared to SiO2 (Al2O3 = 20 nm and SiO2 = 10 nm). The thin layers within the floating gates play a pivotal role in DC bias design. Because their thickness is significantly smaller than the operating wavelength, their influence on the amplitude and phase of the reflected and transmitted waves can be neglected.

They can be fruitfully represented by a physical equivalent circuit in which each patterned graphene between two stratified media (air-quartz or quartz-quartz) is represented as an R-L-C branch (see Supplementary Information C). The supporting substrates can be conceptualized as separate transmission lines, each possessing its own intrinsic impedance and propagation constant, denoted as $\eta_c$=$\eta_0$/$\sqrt{\epsilon_r}$  and $\gamma$ = $\omega \sqrt\epsilon_r/c$, respectively. Here, $\eta_0$ and c refer to the characteristic impedance of free space and speed of light. Utilizing the circuit model, we calculated the amplitude and phase responses for $R_{01}$, $T_0$, absorption, and polarization conversion modes in both reflection and transmission configurations. The determined values for R, L, and C are presented in Supplementary Table S1 for various scenarios. Supplementary Fig. S5(a)–(g) compares the spectra obtained from the circuit model with the spectra from full-wave numerical simulations, demonstrating a excellent agreement in both amplitude and phase responses. These trends are physically understandable and can be thoroughly explained by examining the changes in the real and imaginary parts of graphene conductivity, along with considering the electrostatic capacitive coupling between adjacent elements \cite{71}. Consequently, the development of the circuit model enhances the proposed metasurface design process, and the resulting outcomes closely align with our expectations.

\section{Results and Discussion}
\subsection{control of full space}
\subsubsection{The Scattering Properties of metasurface}
Introducing abrupt phase shifts through coding metasurfaces opens a new avenue for manipulating scattering patterns. Such coded metasurfaces have the ability to reflect or transmit incident waves into anomalous directions, governed by the principles of generalized Snell's law \cite{72}. At the outset, we examine a simple form of coding metasurfaces, characterized by a coding pattern comprising solely two interchangeable coding elements, "0" and "1," yielding opposite reflection phase responses. Different coding sequences of digital particles yield a spectrum of scattering patterns, leading to singular, dual, four, and multiple reflection beams. In order to minimize the EM coupling between adjacent meta-atoms, the proposed metasurface is constructed from M×M unit-cells, that form the so-called super unit cells or lattices. The length of each lattice is ($D_x = D_y = D$), which is equal to PM. Due to the one-to-one connection between different coding patterns of metasurface and their far-field patterns, under normal incidence, the proposed metasurface far-field scattering pattern function can be expressed by \cite{73,74}.

\begin{equation}
    F(\theta,\phi)=f_{m,n} (\theta,\phi) \sum_{m=0}^{M-1} \sum_{n=0}^{M-1} A_{m,n} exp(u)\times exp(v)
\label{eq5}
\end{equation}

Where $u=i m k_0 D sin\theta cos\phi$, $v=i n k_0 D sin\theta sin\phi$, $f_{(m,n)} (\theta,\phi)$ is the pattern function of lattice, $A_{(m,n)}= a_{m,n} exp (i\phi_{m,n}) $ is the complex reflection coefficient, $k_0$ is the free-space wavevector, and $\theta$ and $\phi$ are elevation and azimuth angles. Due to the metasurface units being significantly smaller than the wavelength, During the calculation of the far-field pattern, we can confidently disregard $f_{(m,n)} (\theta,\phi)$. This formula proves advantageous for predicting scattering patterns arising from different coding sequences. Nonetheless, augmenting the overall number of lattices within the metasurface leads to extended computation durations. To circumvent this limitation, a viable solution involves employing a two-dimensional inverse fast Fourier transform (2D-IFFT) from equation \ref{eq5}, resulting in a considerable acceleration of the calculations \cite{23,75}. 

\begin{figure*}
    \centering
    \includegraphics[scale=0.23]{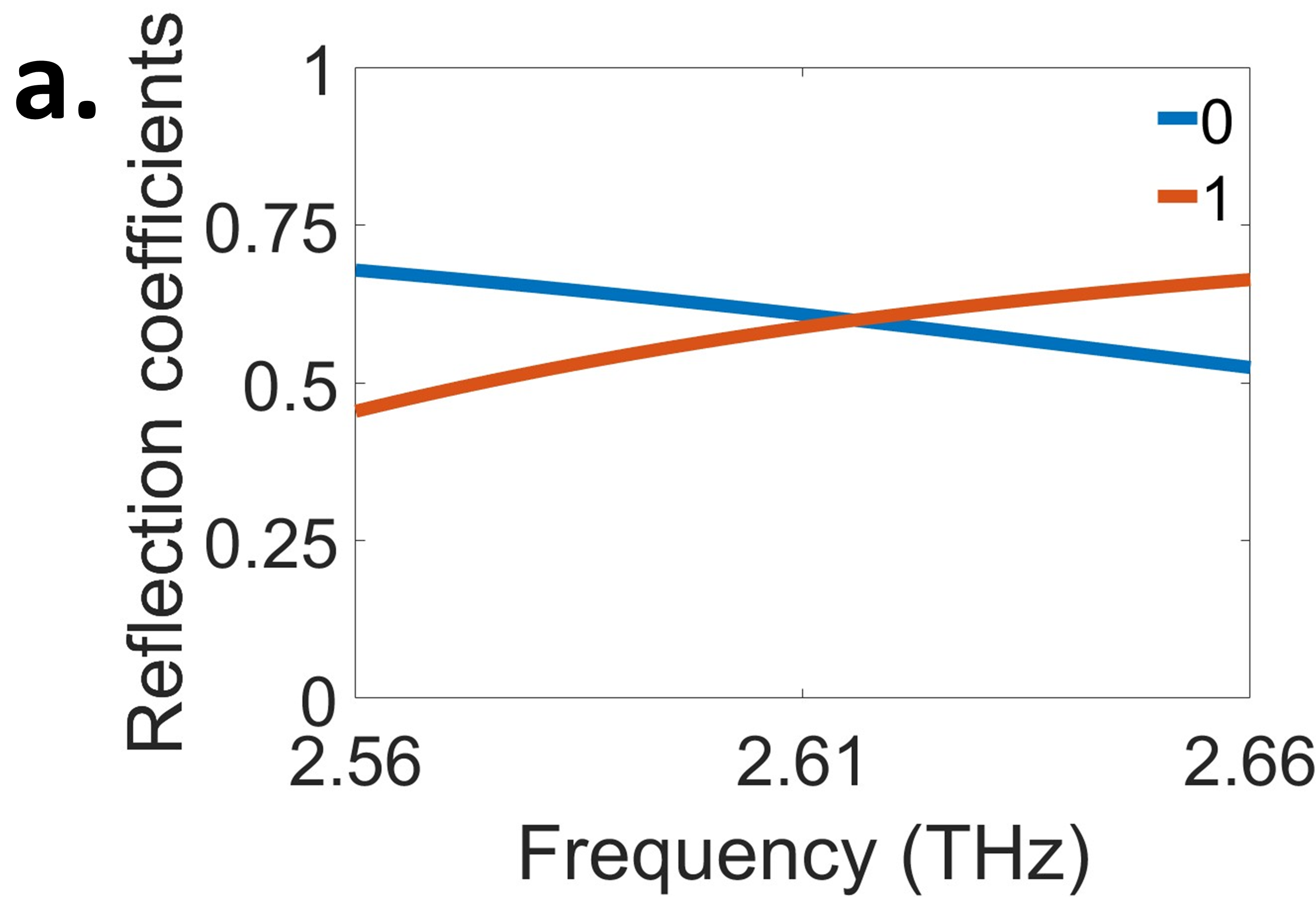}
    \includegraphics[scale=0.23]{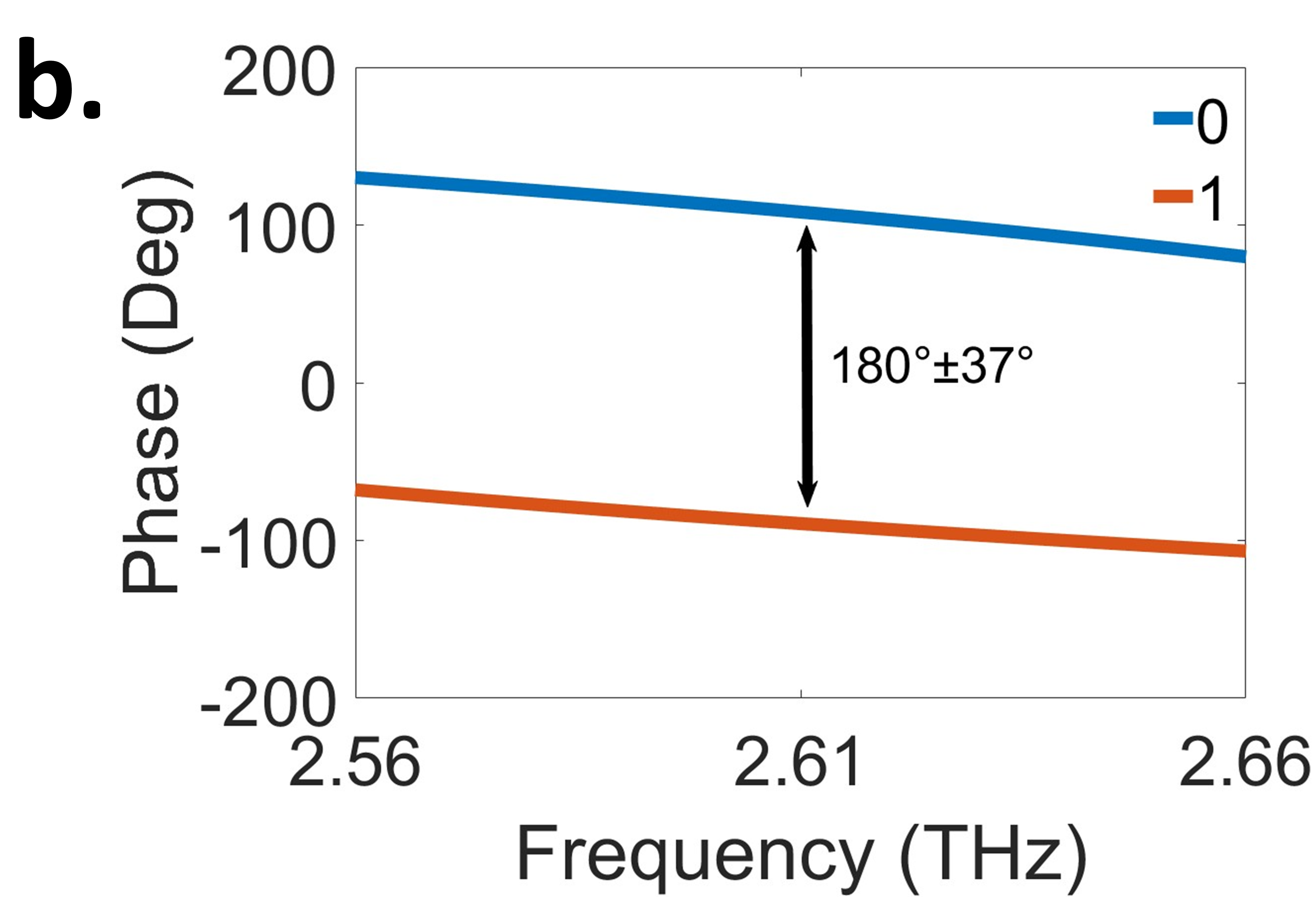}
    \\
    \includegraphics[scale=0.23]{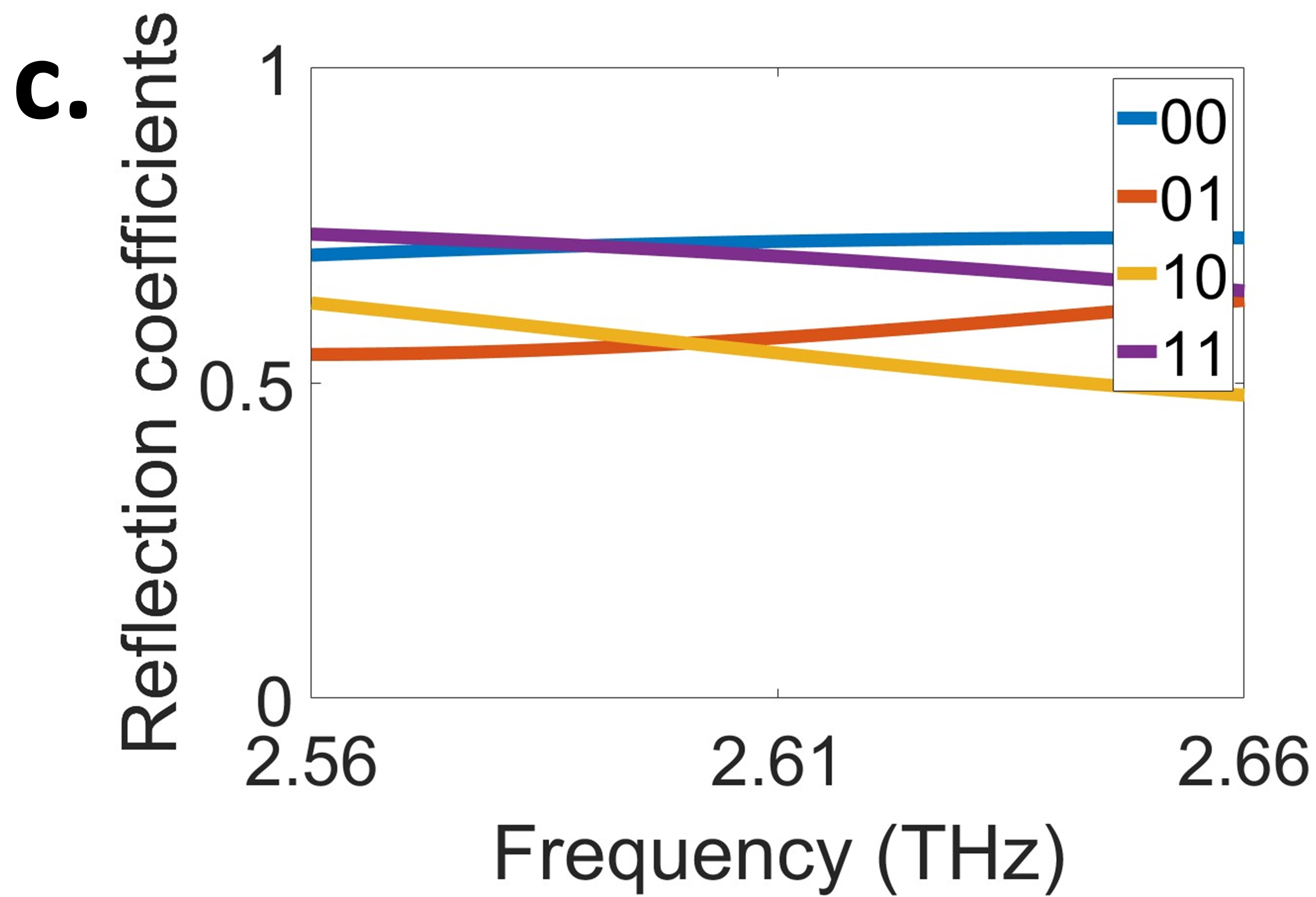}
    \includegraphics[scale=0.23]{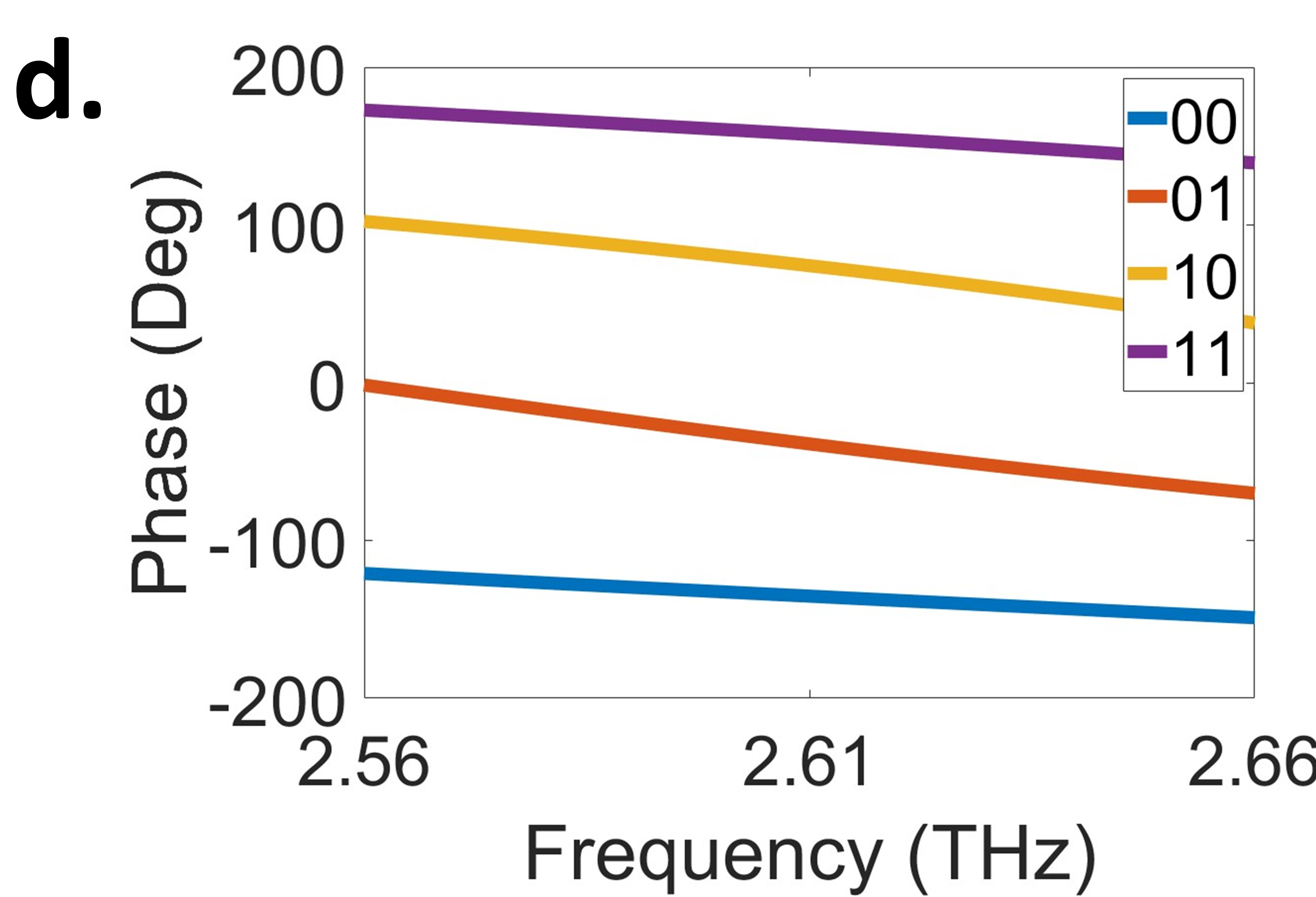}
    \\
    \includegraphics[scale=0.23]{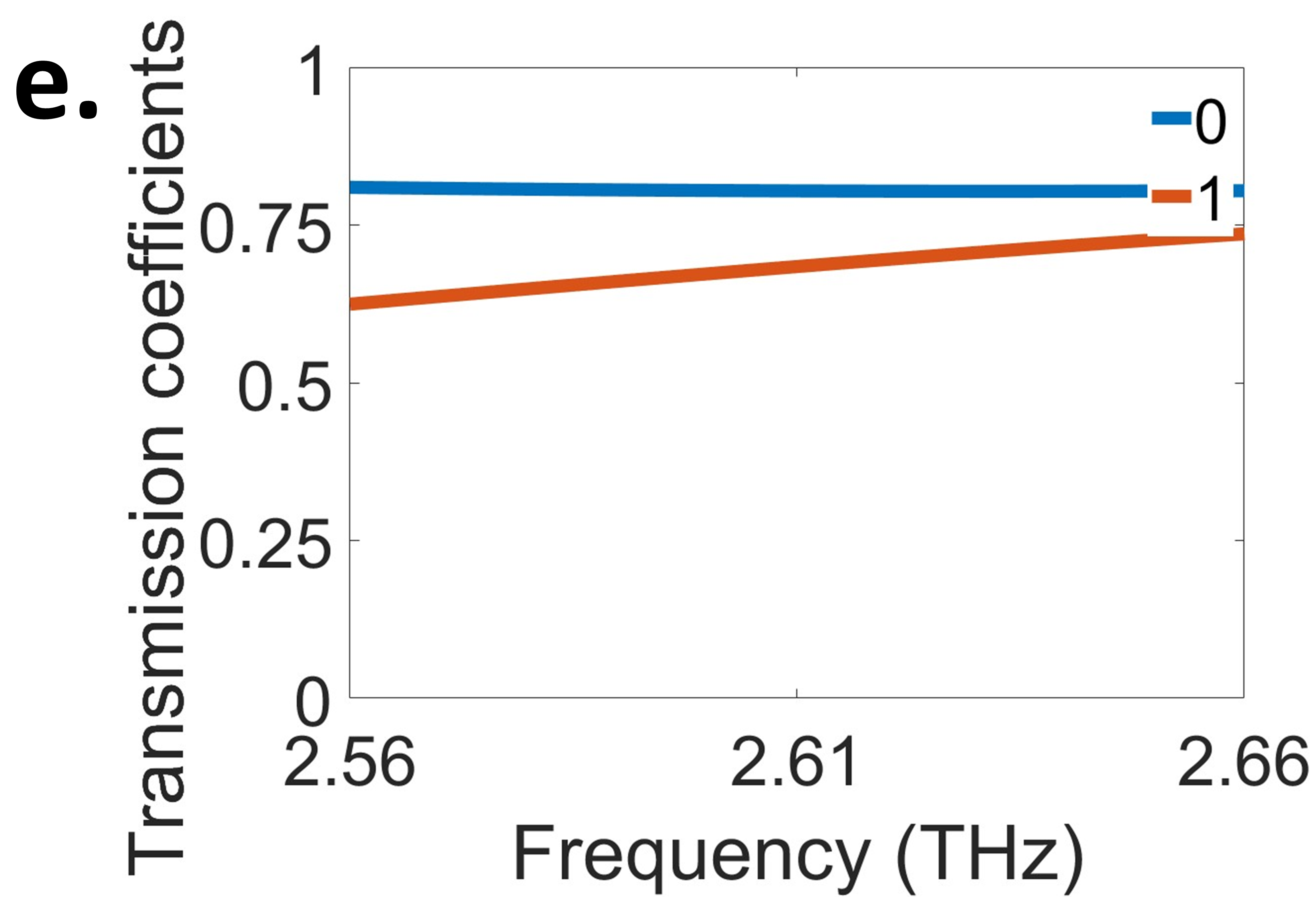}
    \includegraphics[scale=0.23]{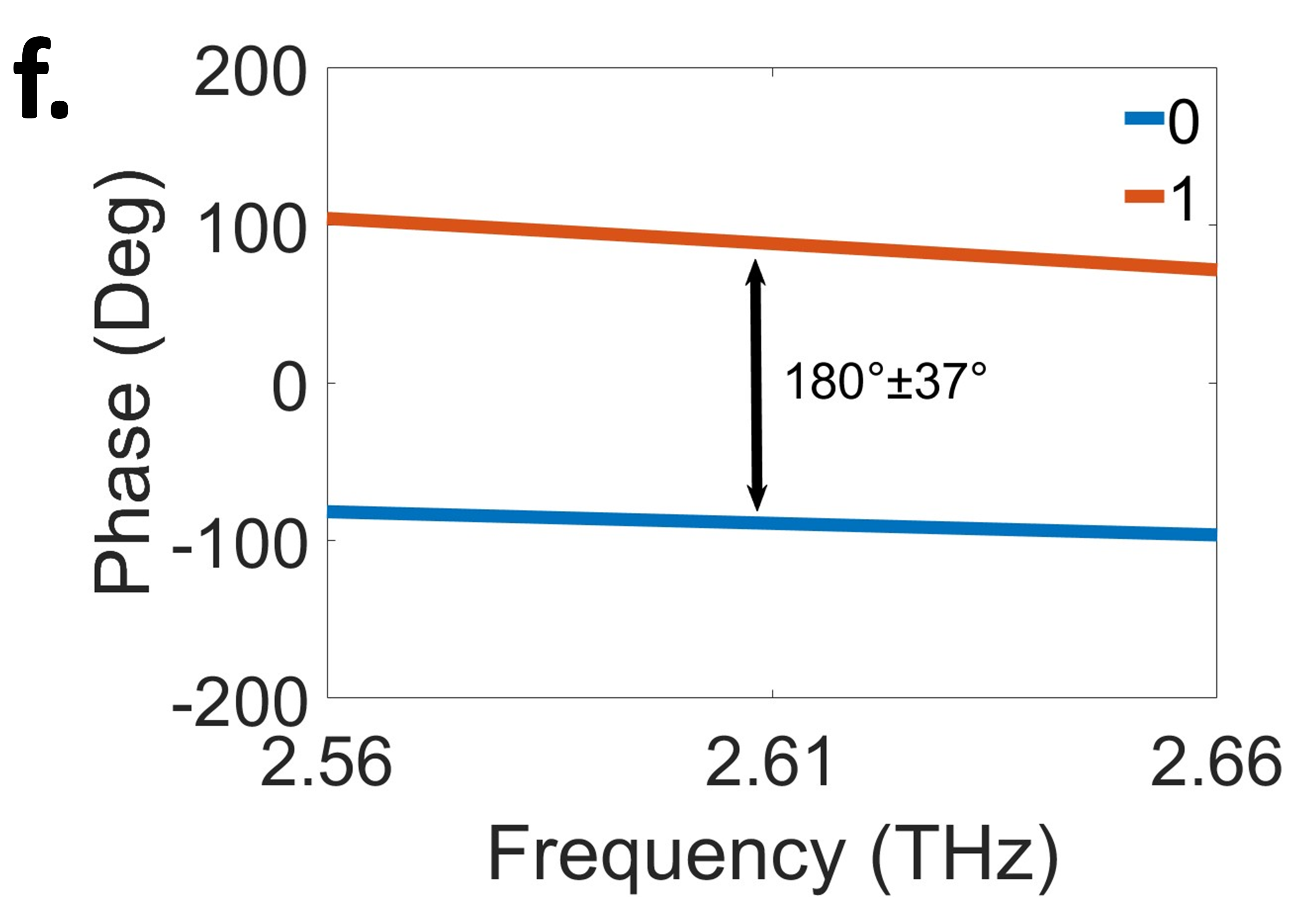}
    \caption{Simulated amplitude for reflection/transmission in 1-Bit mode (a)/(e), reflection/transmission phase in 1-Bit mode (b)/(f), amplitude (c), and reflection phase (d) in 2-Bit mode.}
    \label{FIG:3}
\end{figure*}

\begin{table*}[h]
\centering
\caption{The essential chemical potential for each segment of the graphene meta-atom to govern the wavefront in both reflection and transmission modes.}
\label{Table 1}
\setlength{\arrayrulewidth}{0.4mm}
\scalebox{1.1}{%
\begin{tabular}{cc|clll|ll}
\cline{3-6}
\multicolumn{1}{l}{} & \textbf{} & \multicolumn{4}{c|}{\textbf{G2}} & \multicolumn{1}{c}{\textbf{}} &  \\ \cline{2-7}
\multicolumn{1}{l|}{} & \textbf{G1} & \multicolumn{1}{c|}{0°} & \multicolumn{1}{c|}{90°} & \multicolumn{1}{c|}{180°} & \multicolumn{1}{c|}{270°} & \multicolumn{1}{l|}{\textbf{G3}} &  \\ \hline
\multicolumn{1}{|c|}{} & \multicolumn{1}{l|}{1.5 eV} & \multicolumn{1}{l|}{\cellcolor[HTML]{000000}{\color[HTML]{ECF4FF} 0.5 eV}} & \multicolumn{1}{l|}{} & \multicolumn{1}{l|}{\cellcolor[HTML]{FE0000}{\color[HTML]{FFFFFF} 1.4 eV}} &  & \multicolumn{1}{c|}{0 eV} & \multicolumn{1}{l|}{\textbf{1-bit}} \\ \cline{2-8} 
\multicolumn{1}{|c|}{\multirow{-2}{*}{\textbf{Reflection}}} & \multicolumn{1}{l|}{1.5 eV} & \multicolumn{1}{l|}{\cellcolor[HTML]{000000}{\color[HTML]{FFFFFF} 1.1 eV}} & \multicolumn{1}{l|}{\cellcolor[HTML]{3531FF}{\color[HTML]{FFFFFF} 0.15 eV}} & \multicolumn{1}{l|}{\cellcolor[HTML]{FE0000}{\color[HTML]{FFFFFF} 0.7 eV}} & \cellcolor[HTML]{036400}{\color[HTML]{FFFFFF} 0.4 eV} & \multicolumn{1}{l|}{0 eV} & \multicolumn{1}{l|}{\textbf{2-bit}} \\ \hline
\multicolumn{1}{|c|}{\textbf{Transmission}} & 0 eV & \multicolumn{1}{c|}{\cellcolor[HTML]{000000}{\color[HTML]{FFFFFF} 0 eV}} & \multicolumn{1}{l|}{} & \multicolumn{1}{l|}{\cellcolor[HTML]{FE0000}{\color[HTML]{FFFFFF} 1.5 eV}} &  & \multicolumn{1}{l|}{0 eV} & \multicolumn{1}{l|}{\textbf{1-bit}} \\ \hline
\end{tabular}%
}
\end{table*}

\subsubsection{Graphene-Based Coding design and its circuital representation}
The simulations performed for each metaparticle were performed using the full-wave commercial software, CST Microwave Studio. Periodic boundary conditions are applied in the x- and y-directions while Floquet ports are also assigned to the z-direction. The meta-atoms within the proposed graphene-based metasurface are divided into three distinct components. By tuning the chemical potential of graphene in the first (G1) and second (G2) layers, control of reflected and transmitted waves is attainable real-time, along with manipulation of the EM wavefront. For reflection and transmission modes, control of the phase layer enables the attainment of 1-bit/2-bit and 1-bit metasurface, respectively. In the reflection mode, the chemical potential of G1 is equal to 1.5 eV, causing graphene to behave as a conductor. As a result, the incident wave is reflected through G1. While in the transmission mode, the value of the chemical potential of graphene G1 should be such that graphene acts like a dielectric so that the incident wave passes through the metasurface, so we consider its value to be 0 eV. In both the reflection and transmission modes, wavefront manipulation is achieved by applying appropriate chemical potentials to the graphenes within the second segment (G2). In the reflection mode, the range of graphene's chemical potential within the middle section is adjusted from 0 eV to 1.4 eV to achieve a 1-bit/2-bit phase shift. Conversely, in the transmission mode, the chemical potential is varied within the same section from 0 eV to 1.5 eV to attain a 1-bit phase shift. It's important to highlight that the consideration of a chemical potential exceeding 1.2 is grounded in prior research findings \cite{41,76,77,78}. Of note, the chemical potential of the third segment (G3) remains consistently set at zero in all full-space EM wavefront manipulation scenarios. The chemical potentials applied to the three sections of graphene meta-atoms are denoted as A/B/C. Specifically, A represents the chemical potential of the first layer (G1), B corresponds to the chemical potential of the second layer (G2), and C signifies the chemical potential of the third layer (G3). The comprehensive reflection and transmission spectra, obtained under x-polarized illumination, are illustrated in Fig.\ref{FIG:3} for various sets of chemical potentials applied to the first and second layers. Illustrated in Fig.\ref{FIG:3}(a) and (b) shows, the selection of chemical potentials at 1.5 eV/ 0.5 eV/ 0 eV and 1.5 eV/ 1.4 eV/ 0 eV yields phase responses with a tuned 180-degree phase shift. Remarkably, within the frequency range of 2.56 to 2.66 THz, these responses effectively enable the representation of binary "0" and "1" through a 1-bit encoding scheme. Fig.\ref{FIG:3} (c) and (d) shows the amplitude and phase coefficients for chemical potentials of 1.5 eV/ 1.1 eV/ 0 eV, 1.5 eV/ 0.15 eV/ 0 eV, 1.5 eV/ 0.7 eV/ 0 eV, 1.5 eV/ 0.4 eV/ 0 eV. Phase responses provide a phase shift of 90° from each other for the frequency band 2.55 to 2.65 THz, which can mimic the 2-bit encoding "00", "01", "10", and "11", respectively. Similarly, Fig.\ref{FIG:3} (e) and (f) shows the amplitude and phase transmission coefficients for chemical potentials 0 eV/ 0 eV/ 0 eV and 0 eV/ 1.5 eV/ 0 eV. The phase responses provide a phase shift of 180° from each other for the same frequency band (2.55 - 2.65 THz), which can mimic 1-bit encoding "0" and "1", respectively. The key point is that all meta-atoms coded in two reflection and transmission modes have high reflection and transmission coefficients in the operational bandwidth because they operate far from their resonance frequency. It's worth highlighting that the reflection and transmission phase responses demonstrate a nearly linear trend, a crucial aspect in the design of multi-bit coding meta-atoms.The summary of the above discussion can be found in Table \ref{Table 1}.

\begin{figure*}
    \includegraphics[scale=0.55]{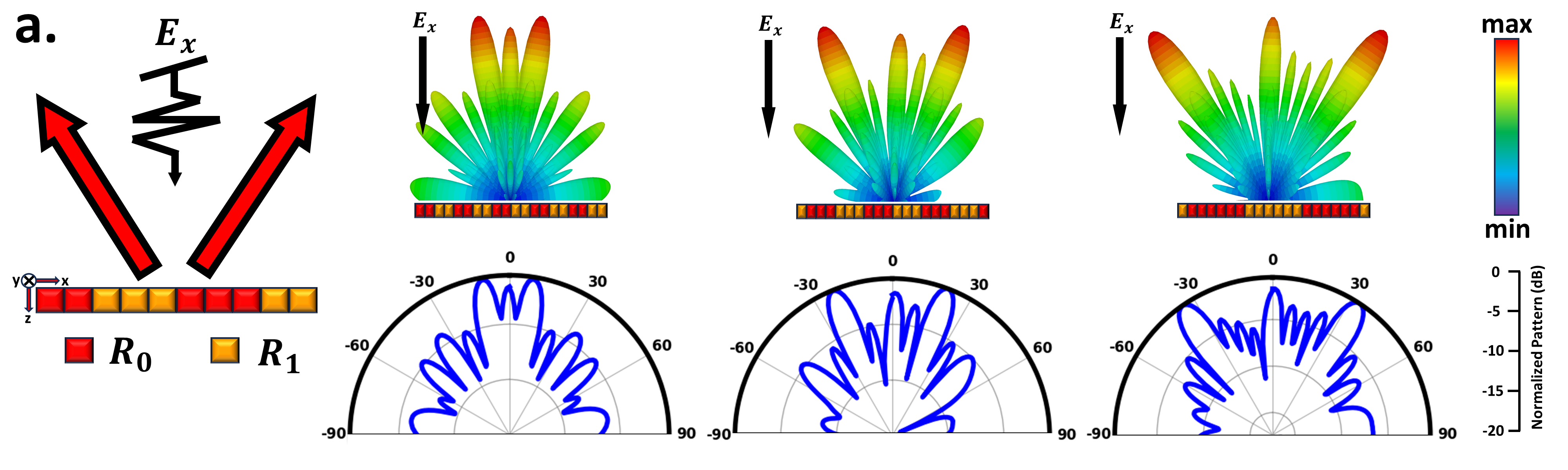}
\\
    \includegraphics[scale=0.55]{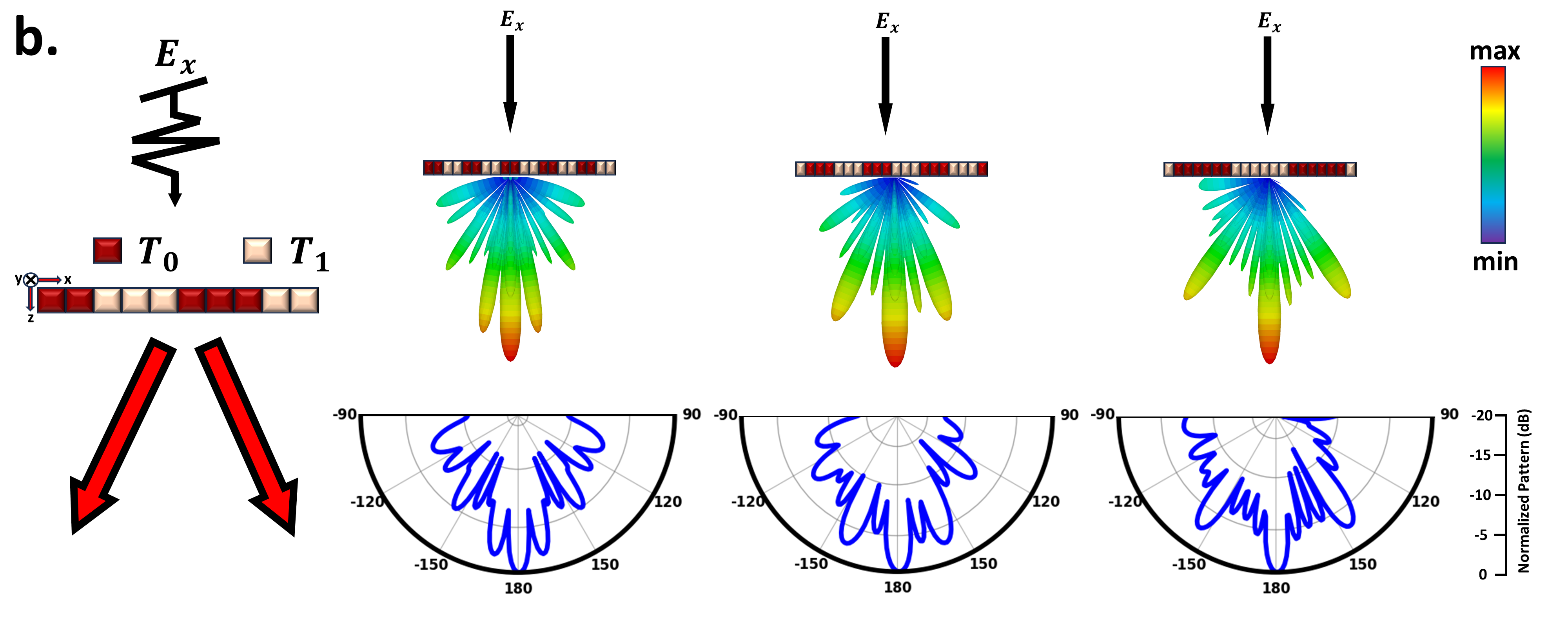}
    \caption{Simulation Outcomes in Reflection and Transmission Modes. (a) Simulation results for the reflection mode using coding sequences "$R_0R_0R_1R_1$...", "$R_0R_0R_0R_1R_1R_1$...", and "$R_0R_0R_0R_0R_0R_0R_1R_1R_1R_1R_1R_1$..." intended to generate twin beams at $\pm$11°, $\pm$22°, and $\pm$34° in the backward half-space, respectively, and (b) Simulation results for the transmission mode using coding sequences "$T_0T_0T_1T_1$...", "$T_0T_0T_0T_1T_1T_1$...", and "$T_0T_0T_0T_0T_0T_0T_1T_1T_1T_1T_1T_1$..." intended to generate twin beams at $\pm$11°, $\pm$22°, and $\pm$34° in the forward half-space, respectively.}
    \label{FIG:4}
\end{figure*}

\subsubsection{Performance of the Reconfigurable Metasurface in 1-bit Reflection and 1-bit Transmission Modes}

As previously mentioned, applying distinct electric field bias values at various locations across the proposed metasurface leads to a switchable coding layout that enables real-time manipulation of EM waves. The intelligent metasurface, comprising 20 × 20 meta-atoms, is designed to govern EM wavefronts on both sides of space. When exposed to a normal x-polarized plane wave propagating along the z-direction, the metasurface reveals the corresponding far-field patterns at f = 2.61 THz.  First, we set the metasurface in reflective mode, which can be equivalently regarded as a reconfigurable reflective metasurface to cover the backward half-space well. In this case, the chemical potential of the reflection-transmission layer graphene is equal to 1.5 eV. It is worth mentioning that abrupt phase changes along the x-direction cause the scattered wave to split into two symmetrical reflection beams and exhibit anomalous reflection behavior \cite{73}. Consequently, dynamic twin-beam production can be achieved by modulating the phase layer's chemical potential and employing stripe coding patterns with a phase difference of 180° consisting of "0" and "1" coding modes. Here, we denote these two states as the coding bytes "$R_{0}$" and "$R_{1}$". The angle direction of the twin beams can be theoretically predicted according to the generalized Snell's law as follows \cite{79}.

\begin{equation}
\theta = \pm arcsin(\frac{\lambda_0}{2\pi} \frac{\Delta\phi}{D})
\label{eq6}
\end{equation}

Where $\lambda_0$ is the wavelength at the central frequency, $\Delta\phi = \pi$ represents the phase difference between lattice, and D = MP (M is the number of meta-atoms in each lattice). As the design examples, we employ three distinct periodic coding sequences along the x-direction within the metasurface and then analyze their resulting reflection scattering patterns. These three sequences are obtained for M = 2, 3, and 6. Hence, the metasurface is theoretically anticipated to deflect the incident x-polarized wave in the reflection mode toward anomalous angles of $\theta$ = $\pm$ 11°, $\pm$ 22°, and $\pm$ 34° within the x–z plane, according to the generalized Snell’s law. To verify the preceding discussion, Fig.\ref{FIG:4}(a) displays the simulated scattering patterns in the reference $\phi$ =0° plane at f = 2.61 THz. The reflected beams split into twin beams symmetrically at the predesigned deviation angles $\theta$ = $\pm$ 11°, $\pm$ 22°, and $\pm$ 34°, demonstrating a good agreement with theoretical predictions. This affirms the dynamic control of dual beams through the the equation \ref{eq6} in the reflection mode. In the transmission state, the chemical potential of the reflection-transmission layer's graphene is equal to 0 eV. In this state, wavefront manipulation is achieved by appropriately biasing the phase layer, effectively encompassing the front half-space. Adopting an approach analogous to that utilized in the reflection mode, we can partition the scattered wave into two symmetric transmission beams, thereby demonstrating an unconventional reflection behavior. Achieving this involves implementing a stripe coding pattern along the x-direction on the phase control layer. Three separate periodic coding sequences are employed for M = 2, 3, and 6 to dynamically control the angles of dual symmetrical beams. This dynamic control of the twin beam is realized through the manipulation of the chemical potential of the phase layers and employing stripe coding patterns along the x-direction. These patterns exhibit a phase difference of 180° and consist of "0" and "1" coding modes. Here, we denote these two states as the coding bytes "$T_0$" and "$T_1$". As shown in Fig.\ref{FIG:4}(b), in the transmission mode, the metasurface deflects the x-polarized wave towards anomalous angles of $\theta$ = $\pm$ 11°, $\pm$ 22°, and $\pm$ 34° within the x-z plane. Notably, these results not only align with the forecasted results from the equation \ref{eq6} but also correspond to the observed functionality in the reflection mode, albeit within a different half-space arrangement. 

\begin{figure*}
    \includegraphics[scale=0.55]{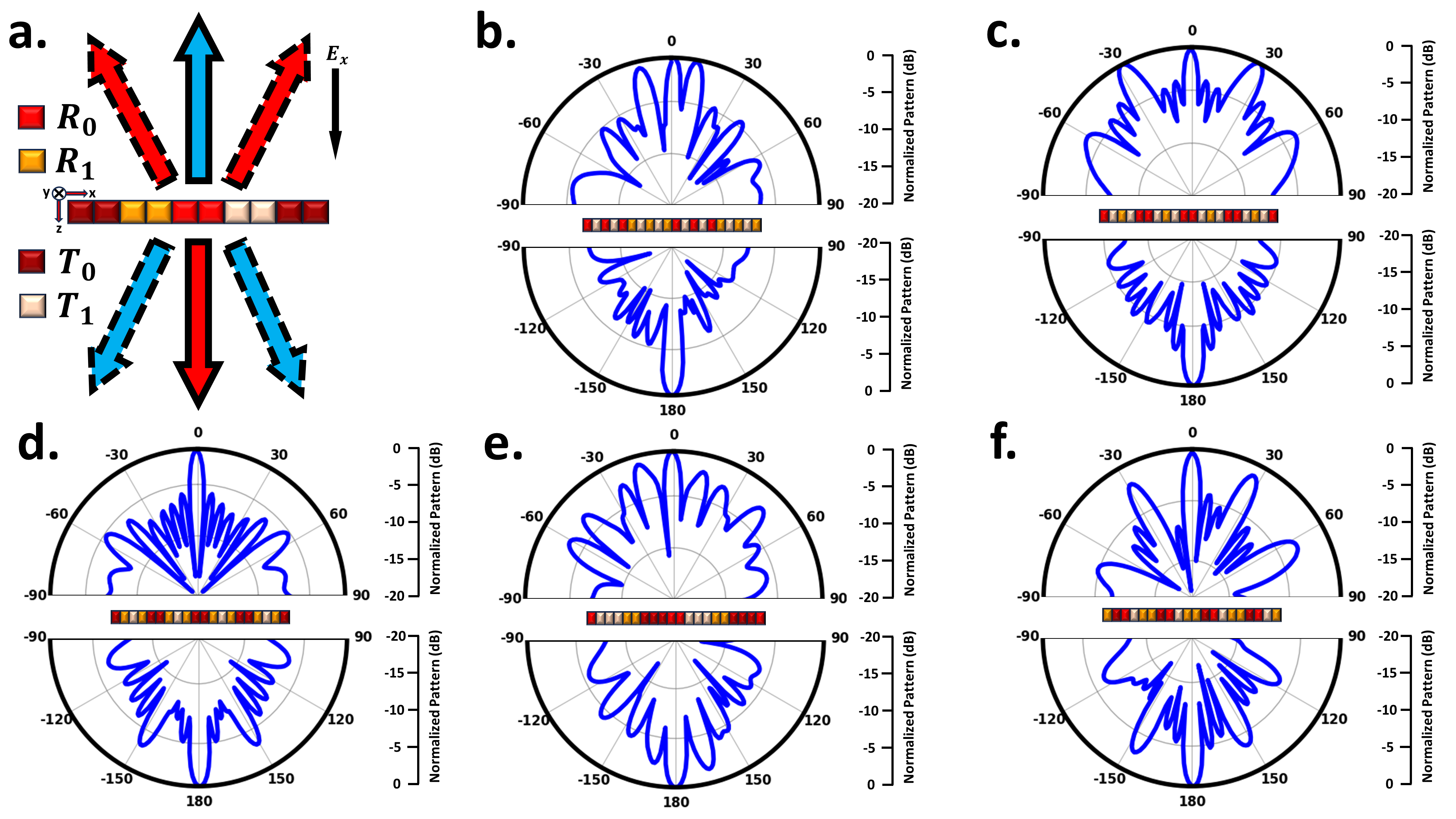}
    \caption{Illustrative Schematic and Simulated Results for Simultaneous 1-bit Reflection/Transmission Mode. (a) Schematic outlining the design concept. The far-field pattern of the entire simulated hypersurface space, which operates concurrently in reflection and transmission modes, with the coding sequence (b) "$R_0T_1R_0T_1R_1T_1R_1T_1R_1$...", (c) "$R_0T_1R_1T_1R_0$...", (d) "$T_0R_1T_1R_1T_0$...", (e) "$R_0T_1T_1T_1R_1R_1T_0T_0T_0R_0$...", and (f) "$R_1T_0R_0T_1R_0$...".}
    \label{FIG:5}
\end{figure*}

To achieve wave control full space, we simultaneously put the metasurface in both reflection mode (backward half-space) and transmission mode (forward half-space). As illustrated in Fig.\ref{FIG:5}, to achieve this objective, we position the lattice associated with reflection and transmission modes adjacently to each other. By establishing different phase functions, we can dynamically manipulate the wavefront in both the reflection and transmission modes. Figure \ref{FIG:5}(a) shows the general schematic of the design for the multifunctional metasurface for the phase distribution in two modes of reflection and transmission. Through simultaneous manipulation of reflection and transmission, we can attain identical or distinct functionalities, exemplified by the control of dual beams in reflection mode-direct propagation transmission mode (indicated by red arrows), direct propagation reflection mode-control of dual beams in transmission mode (indicated by blue arrows), and control of dual beams in reflection mode-control of dual beams in transmission mode within two distinct half-spaces (indicated by red and blue dashed arrows). In the scenario of controlling dual beams in reflection mode-direct propagation transmission, the generation of reflective twin beams is attainable by arranging the digital states "$R_0$" and "$R_1$" in a striped coding pattern, while the direct transmission mode is realized by employing lattices with identical phases ("$T_0$" or "$T_1$"). As depicted in Fig.\ref{FIG:5}(b), the striped arrangement of lattices with a 180° phase difference in the reflection mode results in the transformation of the reflected wave into two symmetrical beams with an angle of $\theta$ = $\pm$ 14°, as previously described. On the other hand, in the transmission mode, where there is no phase difference between the lattices, the incident wave passes through without altering its direction. Through precise manipulation of lattice parameters and arrangement, the angle of the two beams can be effectively modulated to $\theta$ = $\pm$ 28° within the reflection mode. Remarkably, this alteration occurs while preserving the unaltered wave performance within the transmission mode, as demonstrated in Fig.\ref{FIG:5}(c). Illustrated in Fig.\ref{FIG:5}(d), interchanging the coding patterns between reflection and transmission modes induces alterations in the beam patterns across both the backward and forward half-space ultimately yielding direct propagation reflection mode-dual beam control in transmission mode. Evidently, in the reflection mode, the incident wave is reflected without directional alteration, while in the transmission mode is achieved symmetrical twin beams. 
In the dual beam control mode that encompasses both the backward and forward half-spaces, simultaneous manipulation is achieved through stripe arrangement (where lattices maintain a phase difference of 180°) applied to both reflection and transmission modes. This configuration enables the concurrent control of dual beams across both spatial halves. Illustrated in Fig.\ref{FIG:5}(e) and \ref{FIG:5}(f) are the simulated outcomes demonstrating the generation of dual beams at $\theta$ = $\pm$ 14° and $\pm$ 28° angles for both reflection and transmission modes. Furthermore, with the modification of supercell dimensions, diverse angles can be attained, facilitating the control of both spatial halves. This methodology enables the dynamic manipulation of waves across the forward and backward spaces, achieved through the appropriate arrangement and distribution of phase layers and the reflection-transmission control layer.

\begin{figure*}
    \centering
    \includegraphics[scale=0.55]{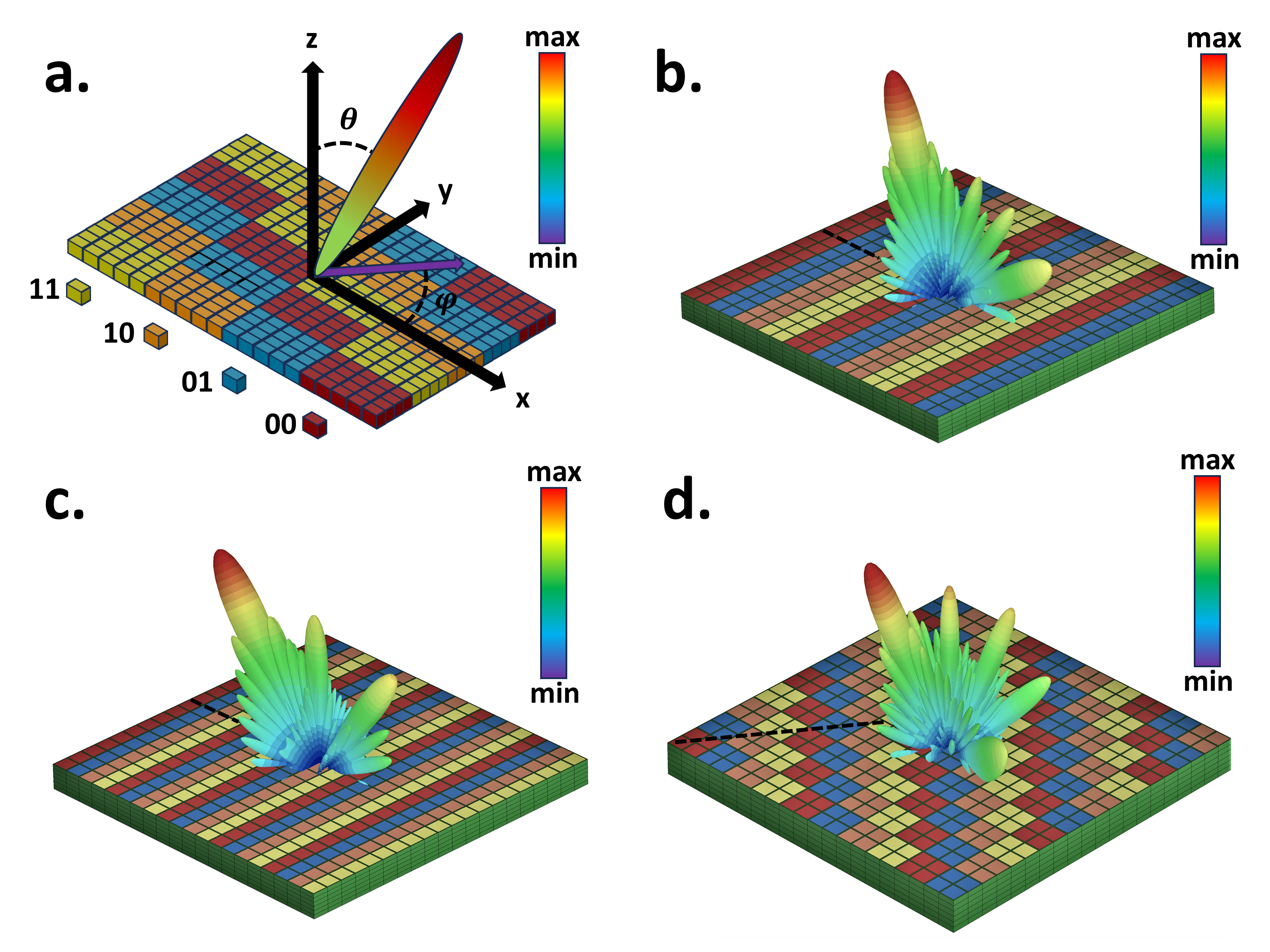}
    \caption{Full-wave simulation results of 3D scattering patterns in 2-bit reflection mode for demonstration of the anomalous reflection into an arbitrary pre-determined direction. (a) The proposed metasurface is suitably programmed with different switchable gradient coding sequences (00 01 10 11 ...) varying along both vertical and horizontal directions.(b) $\theta$ = 16.5°, $\phi$ = 180°, (c) $\theta$ = 35°, $\phi$ = 180° and (d) $\theta$ = 35°, $\phi$ = 225°.}
    \label{FIG:6}
\end{figure*}

\subsubsection{Performance of the Reconfigurable Metasurface in 2-bit Reflection and 1-bit Transmission Modes}

Additionally, owing to the 2-bit (00, 01, 10, 11) phase control in the reflection mode, we explore the generation of reflected beams with arbitrary $\theta$ and $\phi$ angles with the appropriate phase control through the phase control layer. It's crucial to highlight that manipulating reflected or transmitted beams using a 1-bit coding metasurface is not achievable \cite{27}. Phase gradient encoding sequences fabricate synthetic surfaces with the capacity to establish predetermined in-plane wave vectors, thereby allowing precise control over the orientation of the reflected wavefront. Leveraging the lateral phase gradient, a suitably engineered transverse phase discontinuity profile implemented on the metasurface can adeptly steer the incident wave along a new predefined trajectory \cite{32}. according to the generalized Snell's law, the direction of the main rays ($\theta_m$, $\phi_m$) can be written as follows \cite{72}:

\begin{equation}
sin\theta_{m} = \frac{\lambda_0}{2\pi}[(\frac{\Delta\phi_{x}}{D_{x}})^2+(\frac{\Delta\phi_{y}}{D_{y}})^2]^{\frac{1}{2}}
\label{eq7}
\end{equation}

\begin{equation}
tan\phi_{m} = \frac{\Delta\phi_{y}}{\Delta\phi_{x}}\frac{D_{x}}{D_{y}}
\label{eq8}
\end{equation}

Here, $\Delta\phi_{y}$ and $\Delta\phi_{X}$ are the phase differences of lattices along the y and x-directions, respectively and $D_y $ = $D_x $ = MP. Clearly, through design of the coding sequences, it becomes possible to directed the reflected beam to any desired pre-defined angle within each of the four quadrants in the backward half-space. Phase gradient encoding sequences denoted as "$R_{00}$", "$R_{01}$", "$R_{10}$", "$R_{11}$" are designed in both the horizontal and vertical directions to achieve the desired $\theta_m$ and $\phi_m$ angles as depicted in Fig.\ref{FIG:6}(a). To explore the redirection of the reflected wave to a predefined target angle, as can be seen from Fig.\ref{FIG:6}(b), we employ $\Delta\phi_{x} = \pi/2$ and $\Delta\phi_{y}$ = 0°-form phase gradient encoding sequences. These sequences are utilized to reflect radiation rays at specific angles, namely $\theta_m$ = 16.5° and $\phi_m$ = 180°, while the value of M=2. By reducing (increasing) the value of M, the $\theta$ angle can be increased (decreased). As illustrated in Fig.\ref{FIG:6}(c), when M=1, the radiation wave is reflected at an angle of $\theta_m$  = 35° and $\phi_m$ = 180°. In the phase distribution depicted in Fig.\ref{FIG:6}(d), the phase gradient's coding sequence is designed in a manner that $\Delta\phi_{x} = -\pi/2$ and $\Delta\phi_{y} = \pi/2$. Consequently, this phase distribution allows for the reflection of the radiation wave in the direction corresponding to $\theta_m$ = 35° and $\phi_m$ = 225°. It's worth noting that our experimental results align exceptionally well with the corresponding theoretical expectations, demonstrating the accuracy of our approach. By applying an external DC voltage to the graphenes within the phase control (G2) section, controlled by the FPGA, instant access to a single beam at any desired angle in the reflection mode becomes possible.

\begin{figure*}
    \includegraphics[scale=0.55]{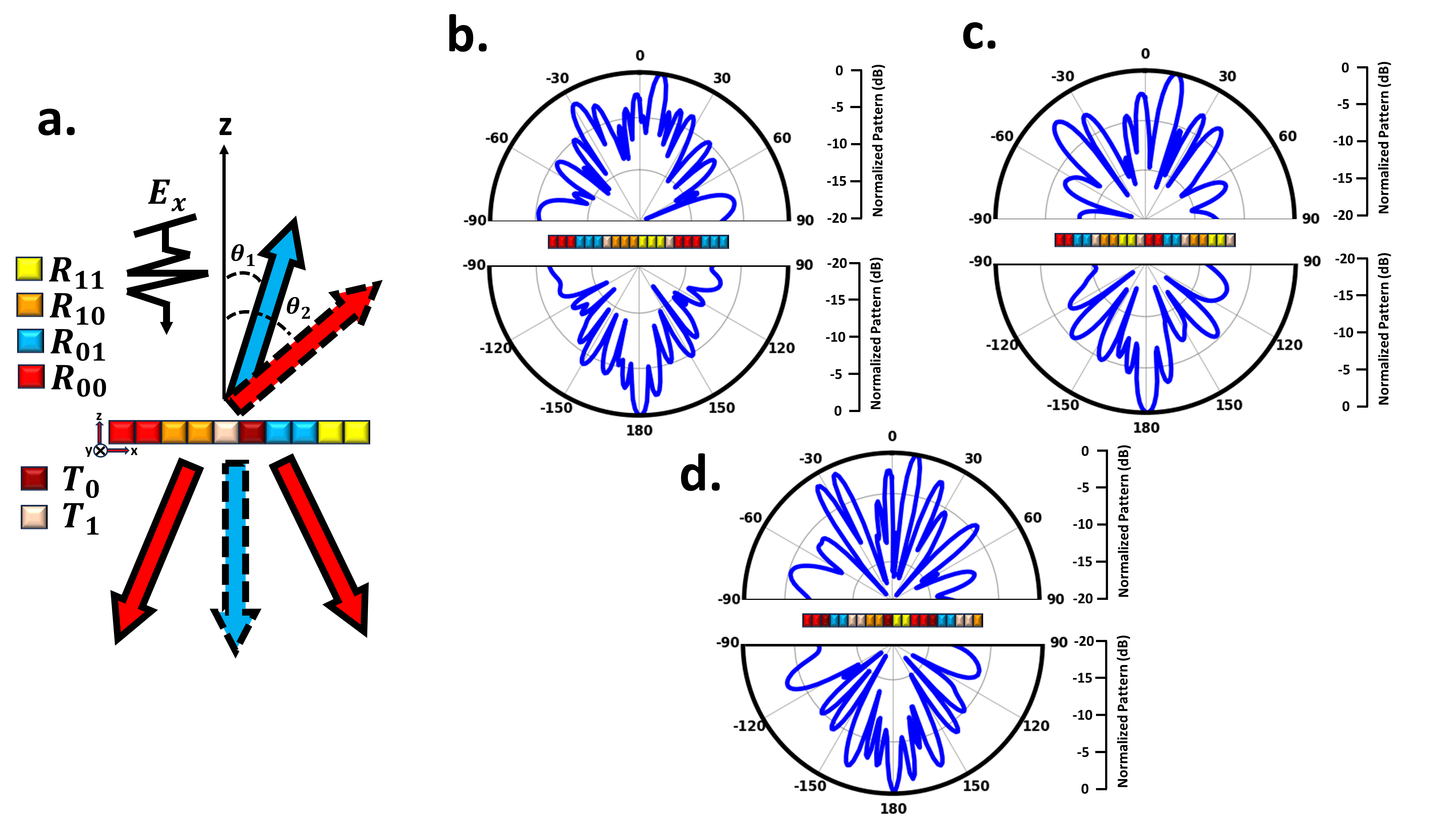}
    \caption{Illustrative Schematic and Simulated Results for Simultaneous 2-bit Reflection/Transmission Mode. (a) Schematic outlining the design concept. The far-field pattern of the entire simulated hypersurface space, which operates concurrently in reflection and transmission modes, with the coding sequence (b) "$R_{00}R_{00}R_{00}R_{01}R_{01}R_{01}T_1R_{10}R_{10}R_{10}R_{11}R_{11}R_{11}T_1$..." (c) "$R_{00}R_{00}R_{01}R_{01}T_1R_{10}R_{10}R_{11}R_{11}$..." (d) "$R_{00}R_{00}T_0R_{01}R_{01}T_1T_1R_{10}R_{10}R_{10}R_{10}T_0R_{11}R_{11}$...".}
    \label{FIG:7}
\end{figure*}

Just like in Fig.\ref{FIG:5}, to attain full space control, we position the meta-atoms responsible for both reflection and transmission modes in proximity each other simultaneously. This deliberate setup, in combination with the generation of an appropriate phase distribution, enables us to manipulate the wavefront in two half-spaces, offering a wide range of functions. These functions encompass not just beam steering in the reflection mode-direct propagation in the transmission mode but also beam steering in the reflection mode-controlling dual beams in the transmission mode. Fig.\ref{FIG:7}(a) illustrates the overall design schematic for the multi-purpose metasurface, featuring a 2-bit phase distribution in the reflection mode and 1-bit in the transmission mode. Through the simultaneous manipulation of reflection and transmission, various functions are achievable. These include beam steering at a specific angle, denoted as $\theta_1$, in the reflection mode-direct propagation in the transmission mode (indicated by blue arrows), beam steering at a specific angle, denoted as $\theta_2$, in the reflection mode-direct propagation in the transmission mode (indicated by red and blue dashed arrows), and beam steering at a specific angle, denoted as $\theta_2$, in the reflection mode-control of dual beams in the transmission mode (indicated by red arrows).In the scenario of beam steering reflective mode-the direct propagation transmission, control of a single beam within the angles $\theta_1$ or $\theta_2$, while phi=180°, can be achieved by configuring the digital modes "$R_{00}$","$R_{01}$", "$R_{10}$" and "$R_{11}$" in a striped coding pattern. Furthermore, the direct transmission mode is established through networks featuring identical phases denoted as "$T_{0}$" or "$T_{1}$".Observing Fig.\ref{FIG:7}(b) and (c), it becomes evident that accomplishing the concurrent beam steering in reflection mode-direct propagation in transmission mode can be attained by arranging the reflection and transmission elements adjacently. Attaining the desired $\theta_1$ and $\theta_2$ angles can  be accomplished through adjustments to the meta-atoms count within the lattice. In reflection mode, organizing lattices into adjacent stripes with a 90° phase offset enables precise control of individual beams. Moreover, when transmission-mode lattices (all composed of $T_1$) are placed alongside reflection mode lattices, we not only achieve wavefront control in the reflection mode but also enable direct wave transmission. It's worth noting that in Fig.\ref{FIG:7}(b), where M=3, and in Fig.\ref{FIG:7}(c), where M=2, we effectively steered the reflected waves at angles of $\theta_1$ =11° and $\theta_2$ =16°, respective. Beyond achieving beam steering in the reflection mode-direct propagation in the transmission mode, we can also gain control over dual beams in the transmission mode. This is accomplished through the precise arrangement of each lattice in transmission mode, where there exists a 180° phase difference between them (T0 and T1), as illustrated in Fig.\ref{FIG:7}(d).

\subsection{control of absorption and polarization}
While achieving real-time wavefront control for both half-spaces represents a groundbreaking accomplishment, we have not confined our metasurface's capabilities to this alone. By appropriate applying chemical potentials to the three sections of graphene, we can further configure the metasurface into two switchable states, that of an absorber and a polarizer. This versatility greatly broadens the potential applications of our proposed metasurface, surpassing conventional expectations. 

\subsubsection{absorption mode}
\begin{figure}
    \includegraphics[scale=0.2]{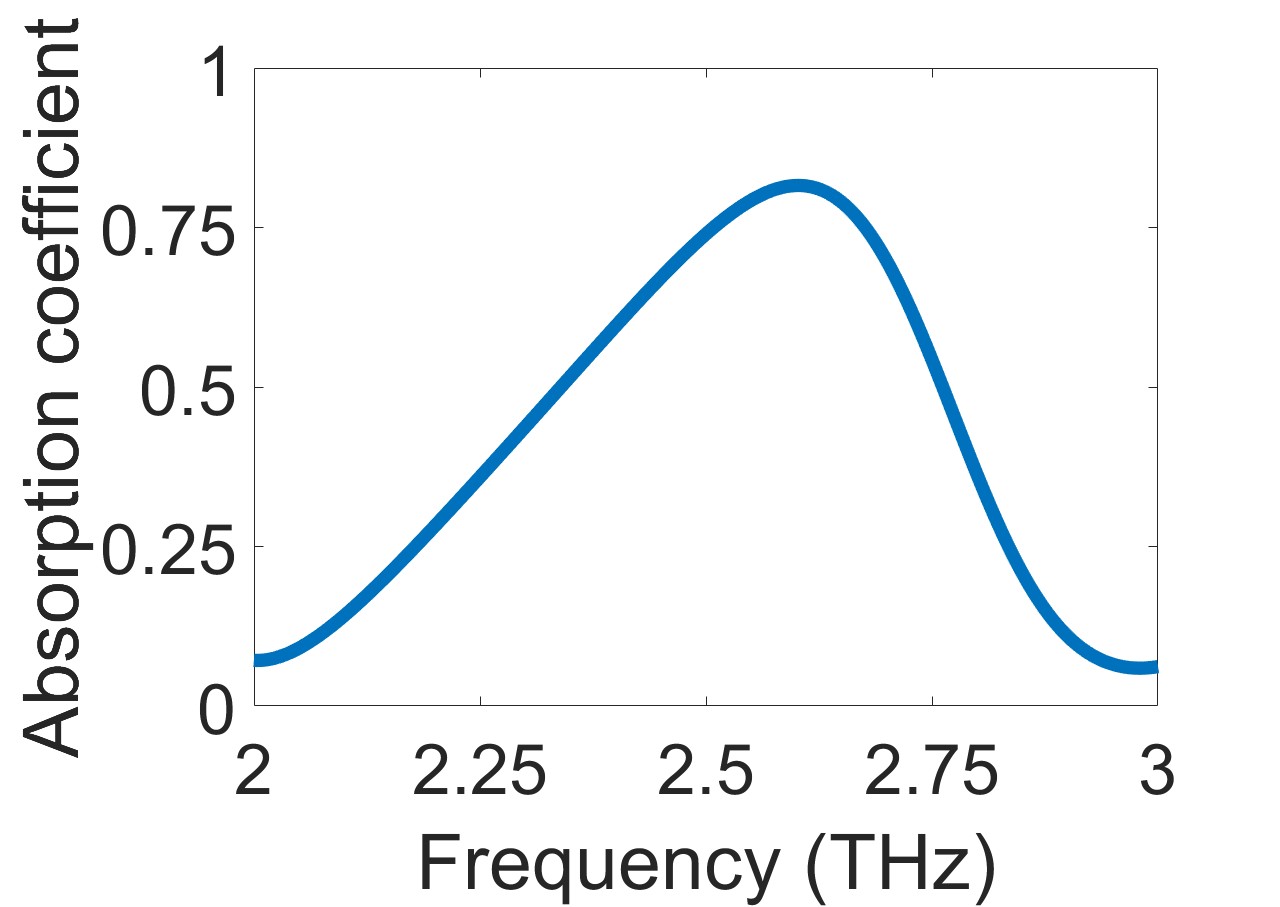}
    \caption{Simulated absorption spectra in the frequency range from 2 to 3 THz.}
    \label{FIG:8}
\end{figure}

The design and development of a metasurface engineered to efficiently absorb EM waves stems from a crucial need in various fields and applications. The concept of EM wave absorption holds immense significance due to its potential to revolutionize several applications such as remote sensing, environmental monitoring, and medical diagnostics. By applying the appropriate chemical potential to the three graphene segments, we can transition the proposed structure into an absorption state. To achieve the desired amplitude of absorption, we deliberately differentiate the relaxation time of third graphene section (G3), setting it to a specific value of 0.1 ps. This is because at $\tau$ = 0.1 ps, the plasmon resonance has reached a state of critical saturation. As the relaxation time increases, a significant portion of the incident light wave is reflected, resulting in a continuous reduction in the electromagnetic wave energy absorbed by the resonance \cite{80}. In the absorption mode, the bottom graphene sheet (G1) is designed to function as an insulator, with its value set at 1.5 eV. We achieve the absorption mode in the desired frequency band (2.56-2.66 THz) by precisely adjusting the chemical potential of the initial graphene sheet (G3). The absorptivity of the structure is calculated by the formula $A(\omega)=1-(R(\omega)+T(\omega))$, where $R(\omega)=|S_{11}|^2$ and $T(\omega)=|S_{21}|^2$ are the reflection and transmission, respectively. $S_{11}$ and $S_{21}$ are the scattering parameters of the reflected and the transmitted wave, respectively. Fig.\ref{FIG:8} demonstrates that by adjusting the chemical potential of the three graphene sections to 1.5 eV / 0 eV/ 0.7 eV, we achieve an absorption amplitude exceeding 0.8. It's worth noting that the performance and results achieved in absorption mode find validation through the equivalent circuit presented in the supplementary materials C. The proposed metasurface not only possesses the capability to absorb EM waves within the desired frequency range (2.56-2.66 THz) but also allows for the fine control of absorption frequencies through the utilization of graphene in the central segment (G2). For a comprehensive understanding of how it can absorb incident waves at different frequencies, please refer to Supplementary Information sections D.

\begin{figure*}
    \centering
    \includegraphics[scale=0.26]{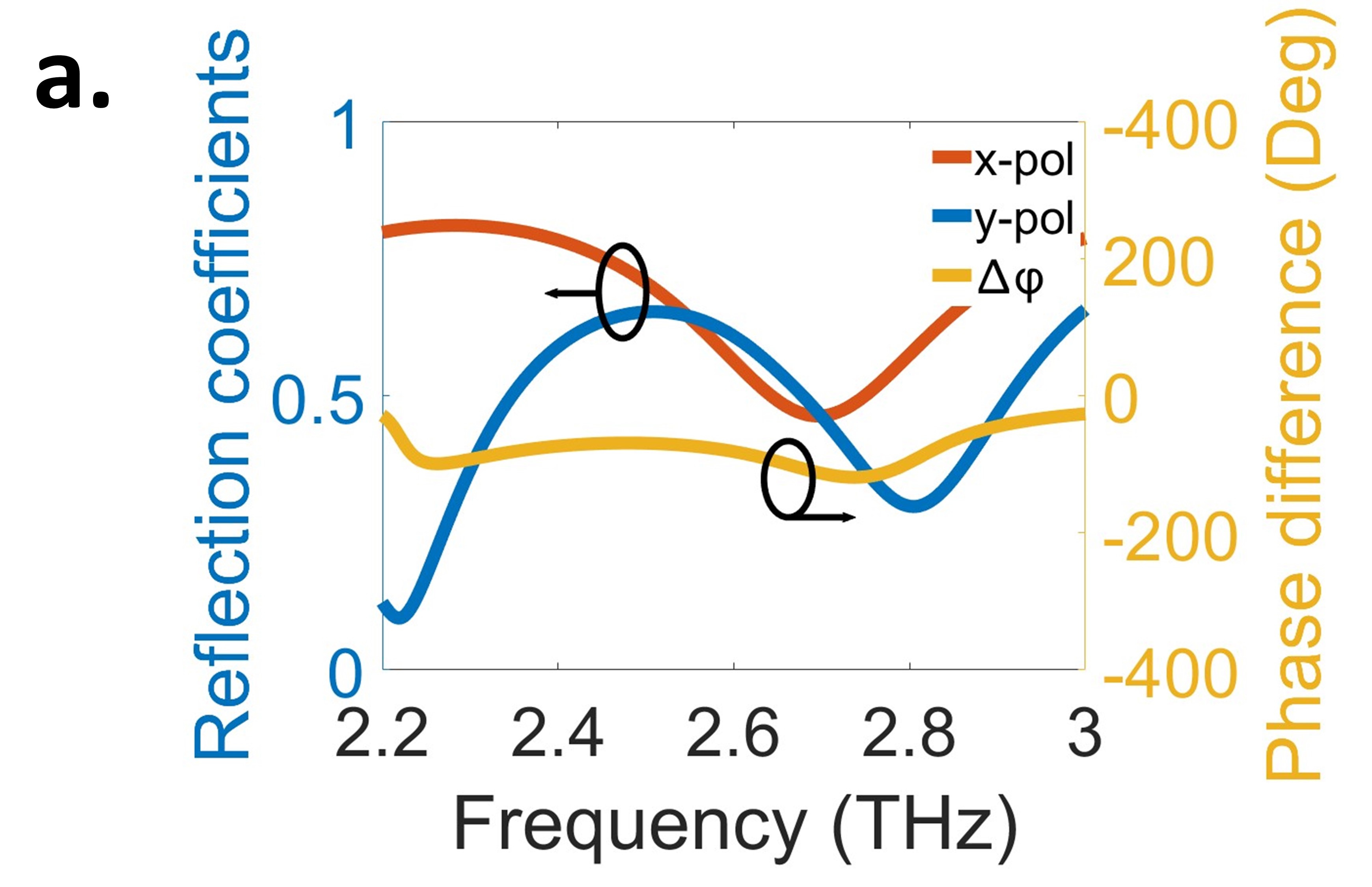}
    \includegraphics[scale=0.26]{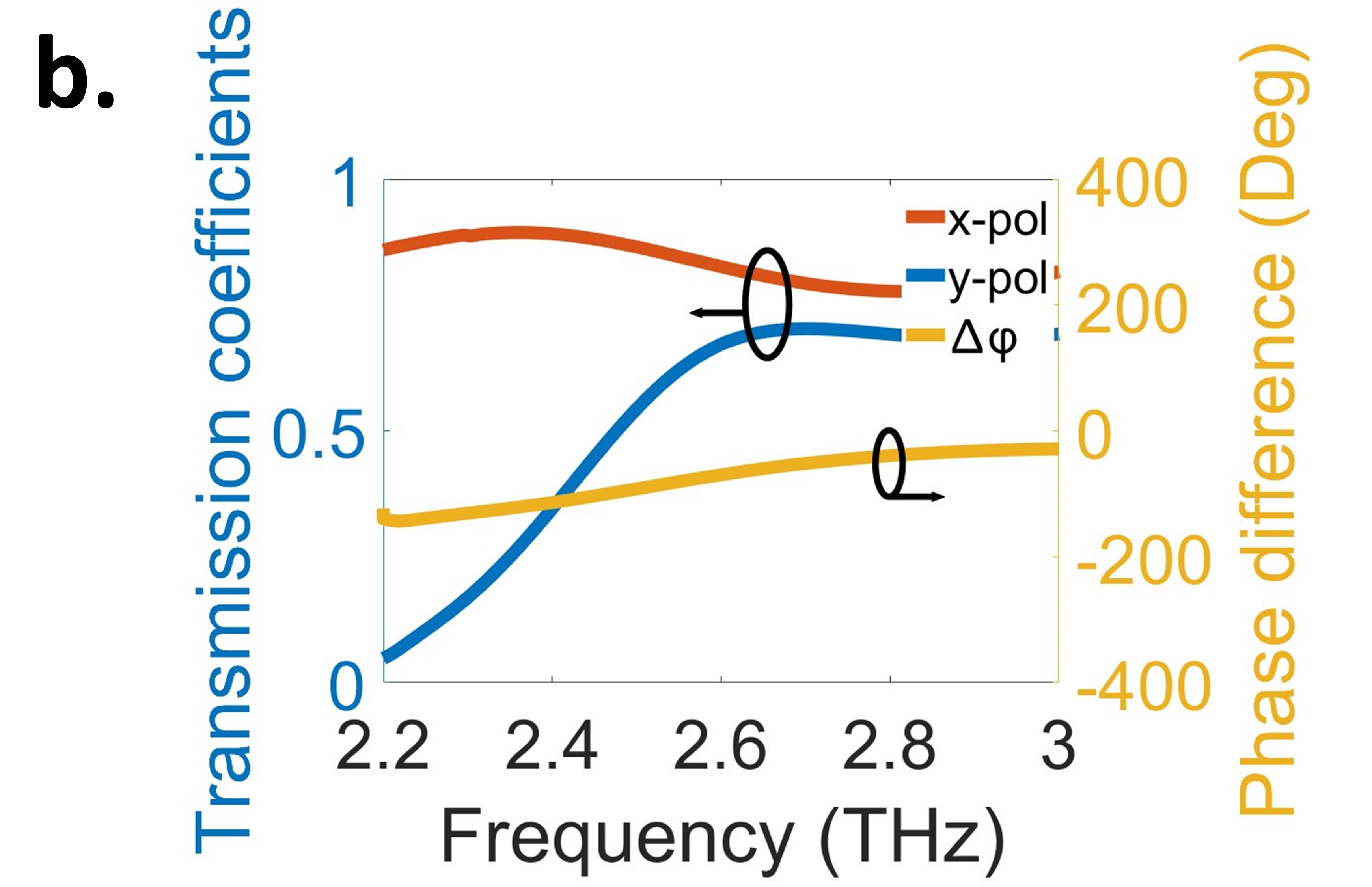}
    \\
    \includegraphics[scale=0.26]{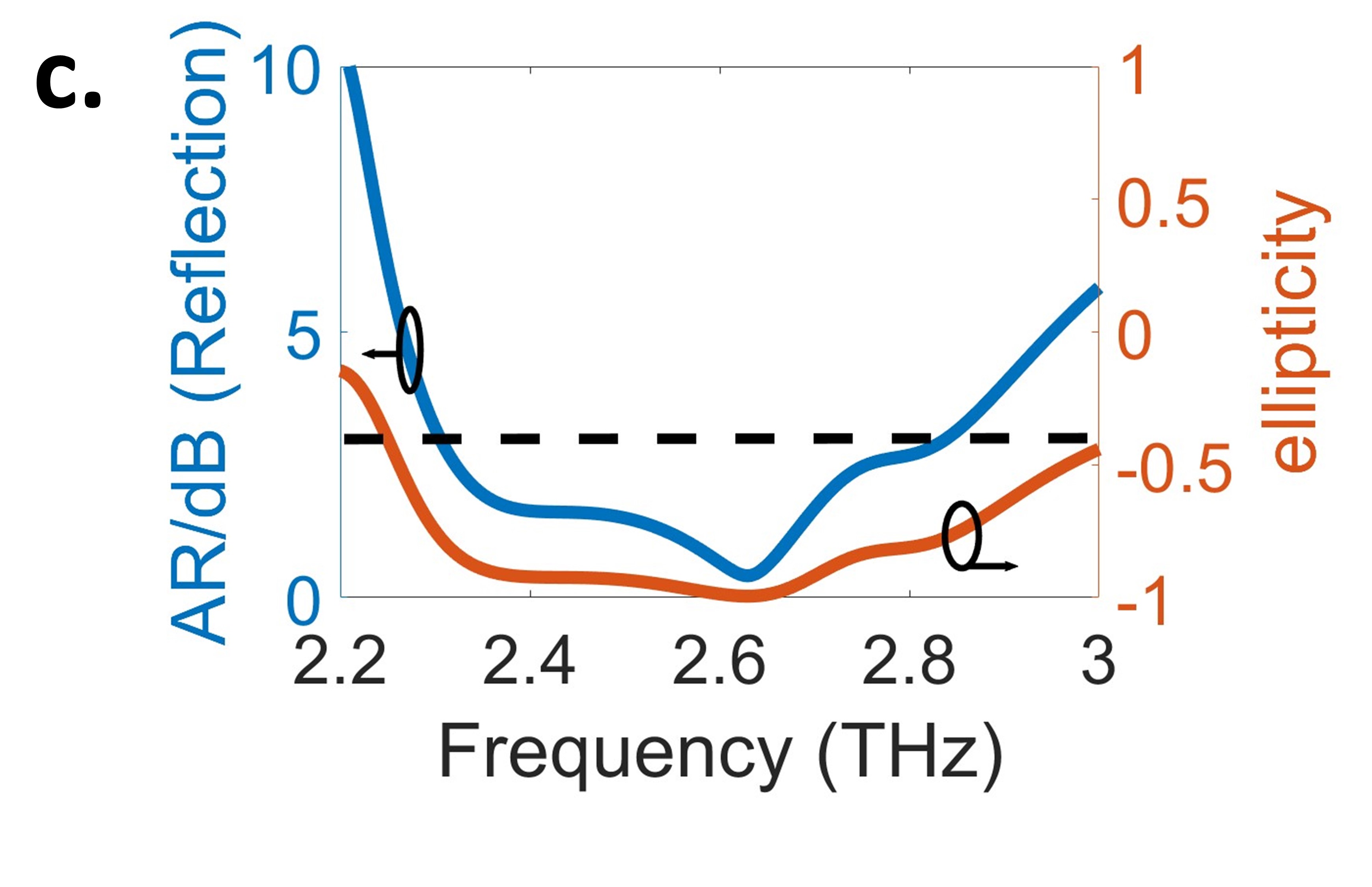}
    \includegraphics[scale=0.26]{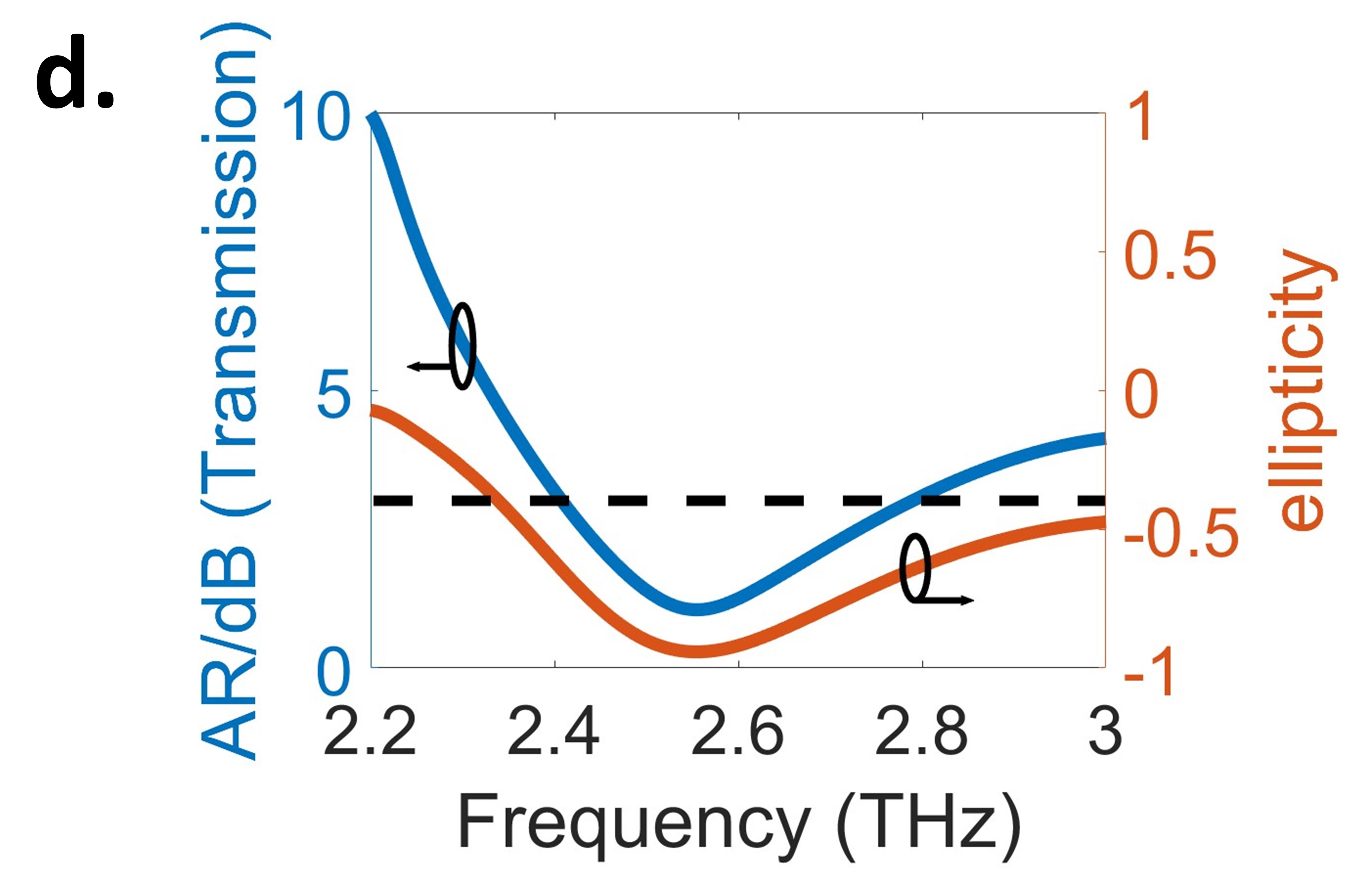}
    \caption{(a) 
Amplitude and phase of the reflected EM wave for x- and y- polarization (b) 
Amplitude and phase of the transmitted EM wave for x- and y- polarization (c) AR and ellipticit of reflected EM wave (d) AR and ellipticit of transmitted EM wave.}
    \label{FIG:9}
\end{figure*}

\subsection{polarization conversion}
Considering that each polarization can independently convey information in communication systems, polarization emerges as a vital and distinctive characteristic of EM waves. Manipulating polarization is of paramount importance in various applications, including telecommunications, radar systems, encryption and optics. Beyond wavefront manipulation on both sides of space and efficient incident wave absorption, we possess the capability for polarization control in both reflection and transmission modes in real-time. By altering the chemical potential in distinct sections of graphene, we can transform incident waves with an x-polarization into circular polarization in both reflection and transmission modes in real time. By applying the appropriate chemical potential to the middle part (G2) and by switching the chemical potential of the first part (G1) to 0 eV and 1.5 eV, we can achieve the conversion of linear polarization to circular polarization in both transmission and reflection modes, respectively. Polarization conversion, in both reflection and transmission modes, can be exclusively accomplished by applying voltage to the two graphene components, G1 and G2, while keeping the chemical potential for G3 = 0 eV. According to the supplementary equation s7, it is evident that to achieve a circularly polarized wave, the reflected or transmitted wave must consist of both x- and y-components. Ideally, these components should possess equal magnitudes $(|E_{xr}|$ = $|E_{yr}|)$, and their phase difference should be either 90° or 270° (-90°), meaning $\Delta\phi$ = 2n$\pi$ $\pm$ $\pi/2$, where n represents an integer. As Fig.\ref{FIG:9}(a) illustrates, for the chemical potential of 1.5 eV/ 0.4 eV/ 0 eV, the reflection coefficient for x- and y-polarization are equivalent, and their phase difference is approximately 90°. In Fig.\ref{FIG:9} (b), for the chemical potential of 0 eV/ 0.5 eV/ 0 eV, the transmission coefficient for x- and y-polarization are also equal to each other and have an approximate phase difference of 90°. We explored the amplitude and phase of both x- and y-polarizations in both reflection and transmission modes using the transmission lines introduced in Supplementary Information c. The results obtained from the circuit model closely match those from the full-wave simulation.

To describe the performance of linear-to-circular polarization conversion in both reflection and transmission modes, we compute the axial ratio (AR) for the reflected and transmitted waves using the formula provided in Eq. (9) \cite{81}:

\begin{equation}
AR = \Bigg(\frac{|R_{xx}^2|+|R_{yy}^2|+\sqrt{a}}{|R_{xx}^2|+|R_{yy}^2|-\sqrt{a}}\Bigg)
\label{eq9}
\end{equation}
Where a = $|R_{xx}^4|+|R_{yy}^4|+2|R_{xx}^4||R_{yy}^2|cos(2\Delta\phi)$, $R_{xx}$ ($T_{xx}$) represents the reflection (transmission) coefficient for co-polarized waves when the incident wave is x-polarized, and $R_{yy}$ ($T_{yy}$) represents the reflection (transmission) coefficient for co-polarized waves when the incident wave is y-polarized. Ideally, in linear-to-circular polarization conversion, the axial ratio should be less than or equal to 3 dB, i.e., AR $\leq$ 3 dB. As evident from Fig.\ref{FIG:9}(c) and (d), the axial ratio for both reflection and transmission modes is below 3 dB. We employ the Stokes parameters, defined as follows, to analyze right-hand and left-hand circular polarization \cite{82}. 

\begin{align}
S_{0} = R_{xx}^2 + R_{yy}^2
\\
S_{1} = R_{xx}^2 - R_{yy}^2
\\
S_{2} = 2R_{xx}^2R_{yy}^2cos(\Delta\phi) 
\\
S_{3} = 2R_{xx}^2R_{yy}^2sin(\Delta\phi)
\label{eq10}
\end{align}

Consequently, real-time conversion of linear to circular polarization is achievable in both reflection and transmission modes. We introduce the concept of normalized ellipticity, denoted as $e = S_{3}/S_{0}$, to characterize circular polarization. When e = 1, it signifies that the reflected or transmitted wave exhibits left-hand circular polarization (LHCP), whereas when e = -1, it indicates that the wave demonstrates right-hand circular polarization (RHCP). As depicted in Fig.\ref{FIG:9}(c) and (d), In both reflection and transmission mode, the reflected and transmitted EM wave is RHCP. The functionality and efficiency of polarization conversion can be assessed by analyzing the Polarization Extinction Ratio (PER), which is elaborated on in Supplementary Information E.

\section{fabrication method}
Initially, the p-Si wafer orientations underwent standard RCA cleaning procedures. Subsequently, to facilitate the growth of ultrathin, high-quality 
SiO2 tunnel oxide, the cleaned p-Si wafers were introduced into a Rapid Thermal Oxidation process. In the third step, the SiO2/p-Si/SiO2 wafers were subjected to a backside buffer oxide etch. Next, the Quartz layer can be applied to a p-Si/SiO2 wafer using a spin-coating solution. 
\\
All sections of graphene used were grown on copper foil (from Graphene Supermarket), and the quartz/Si/SiO2 substrate is the target substrate on which the devices are fabricated. To transfer graphene grown on copper foil to the Si/SiO2 substrate, we employed the wet transfer method as our chosen technique \cite{83}. This approach utilizes PMMA as a supporting layer. A PMMA layer is spun on top of the whole sheet of monolayer graphene grown on Cu foil. Because the CVD process leads to the growth of graphene on both the top and bottom sides of the Cu foil, it is necessary to remove the graphene on the uncoated bottom side of the Cu foil using reactive ion etching (RIE). The PMMA-coated graphene, specifically on the top side of the Cu foil, is subsequently cut into the desired size and then floated on top of a 0.1 M solution of ammonia persulfate (APS). Thanks to the excellent transparency of PMMA, you can visually inspect it directly to determine whether the underlying Cu foil has been adequately removed or not. Next, the graphene covered with PMMA is carefully transferred onto the Si/SiO2 substrate. After allowing the moisture between the Si/SiO2 substrate and graphene to thoroughly evaporate in the air, the PMMA layer is gently removed using warm acetone. Finally, it is transferred onto a spacer consisting of a 20 nm-thick insulating layer (Al2O3) that was uniformly deposited using atomic layer deposition on an ultra-high resistivity p-Si wafer, resulting in the final structure depicted in Fig.\ref{FIG:2}.

\section{conclusion}
In summary, for the first time, we have designed a graphene-assisted reprogrammable intelligent metasurface in the terahertz band, which offers an unprecedented ability to independently and simultaneously manipulate EM waves in real time in both half-spaces. The proposed metasurface possesses the unique capability to independently and simultaneously control wavefronts in both reflection and transmission modes, all within the same polarization and frequency channel. The meta-atom design incorporates three graphene sections. Through the control of each graphene section's chemical potential using an external electronic source, we can manipulate the EM wave to serve various functions. In addition to real-time amplitude control for both reflection and transmission modes, we can also harness graphene to manage the phase of the EM wave. Such capabilities enable us to achieve diverse functionalities, including controlling beam steering and two beams in reflection mode, control of two beam in transmission mode, and combining functionalities from both reflection and transmission modes. Beyond wave control in both half-spaces, our research extended to the manipulation of absorption and polarization states. By altering the chemical potential of each graphene section, we can absorb incident waves across various frequencies. Additionally, by manipulating the polarization of the EM wave, we can achieve the transition from linear to circular polarization in both reflection and transmission modes. The proposed metasurface is expected to have wide applications in such as imaging systems, encryption, remote sensing, miniaturized systems, and next-generation wireless intelligent communications.


%





\ifCLASSOPTIONcaptionsoff
  \newpage
\fi





\bibliographystyle{IEEEtran}
\bibliography{IEEEabrv,Bibliography}

\vfill


\end{document}